\documentclass[11pt]{article}

\usepackage{booktabs}
\usepackage[english]{babel}
\usepackage{amsmath,amssymb,amsbsy,amstext, amsthm, simplewick}
\usepackage{hyperref}
\usepackage{graphicx}
\usepackage{amsfonts}
\usepackage{amssymb}
\usepackage[small]{caption}
\usepackage{upgreek}
\usepackage{cite}

\usepackage{colortbl}
\definecolor{lightgreen}{cmyk}{0.2, 0, 0.2, 0.2}
\definecolor{lightgray}{cmyk}{0.1,0.2,0,0.1}
\definecolor{lightgray2}{cmyk}{0.1,0.1,0,0.1}

\makeatletter
\newlength{\apb@width}
\newcommand{\autoparbox}[2][c]{\settowidth{\apb@width}{#2}\parbox[#1]{\apb@width}{#2}}

\makeatother

\newcommand{\gsim}{\mathrel{\lower.8ex\vbox{\lineskip=.1ex\baselineskip=0ex \hbox{$>$}\hbox{$\sim$}}}}
\newcommand{\lsim}{\mathrel{\lower.8ex\vbox{\lineskip=.1ex\baselineskip=0ex \hbox{$<$}\hbox{$\sim$}}}}

\newcommand{\hpi}{\tilde{\pi}}
\newcommand{\hpiS}{$\tilde{\pi}$-short}
\newcommand{\hpiL}{$\tilde{\pi}$-long}
\newcommand{\hpiSx}{$\tilde{\pi}$-short }
\newcommand{\hpiLx}{$\tilde{\pi}$-long }
\newcommand{\hpiN}{\tilde{\pi}^{0}}
\newcommand{\fpi}{f_{\tilde{\pi}}}
\newcommand{\hrho}{\tilde{\rho}}

\newcommand{\SU}{\text{SU}}
\newcommand{\U}{\text{U}}

\numberwithin{equation}{section}

\def\beq{\begin{equation}}
\def\eeq{\end{equation}}

\def\bea{\begin{eqnarray}}
\def\eea{\end{eqnarray}}

\def\beq{\begin{equation}}
\def\eeq{\end{equation}}
\def\bea{\begin{eqnarray}}
\def\eea{\end{eqnarray}}

\DeclareRobustCommand{\SkipTocEntry}[4]{}

\begin{document}

\begin{titlepage}

\setcounter{page}{1} \baselineskip=15.5pt \thispagestyle{empty}
\begin{flushright}UTTG-28-15
\end{flushright}

\bigskip
\vspace{1.2cm}
\begin{center}
{\fontsize{18}{32}\selectfont  \bf Shedding light on diphoton resonances}
\end{center}

\vspace{0.5cm}
\begin{center}
{\fontsize{14}{30}\selectfont   Nathaniel Craig$^{\spadesuit}$, Patrick Draper$^{\spadesuit}$, Can Kilic$^{\blacklozenge}$, and Scott Thomas$^{\clubsuit}$}
\end{center}

\vspace{0.2cm}

\begin{center}
\vskip 8pt
\textsl{$^\spadesuit$ Department of Physics, University of California, Santa Barbara, CA 93106, USA}
\vskip 7pt
\textsl{$^ \blacklozenge$ Theory Group, Department of Physics and Texas Cosmology Center, \\
The University of Texas at Austin, Austin, TX 78712, USA}
\vskip 7pt
\textsl{$^\clubsuit$ New High Energy Theory Center, Rutgers University, Piscataway, NJ 08854, USA}
\end{center}

\vspace{1.2cm}
\hrule \vspace{0.3cm}
{ \noindent {\sffamily \bfseries Abstract} \\[0.1cm]

The experimental and theoretical implications of heavy digauge boson resonances 
that couple to, or are comprised of, new charged and 
strongly interacting matter are investigated. 
Observation and measurement of ratios of the resonant digauge boson channels 
$WW$, $ZZ$, $\gamma \gamma$, $Z \gamma$, and $gg$ in the form of dijets, 
provide a rather direct -- and for some ratios a rather robust -- probe of the gauge representations of the 
new matter.  For a spin-zero resonance with the quantum numbers of the vacuum,
the ratios of resonant $WW$ and $ZZ$ to $\gamma \gamma$ channels, 
as well as the longitudinal versus transverse polarization fractions in the $WW$ and $ZZ$ 
channels, provide probes for possible mixing with the Higgs boson, 
while di-Higgs and ditop resonant channels, $hh$ and $tt$, provide somewhat less sensitivity. 
We present a survey of possible underlying models for digauge boson resonances by considering 
various limits for the mass of the new charged and strongly interacting 
matter fields as well as the confinement scale of new hypergauge interactions under which they 
may also be charged.  In these limits, resonances may be included as elementary 
weakly coupled spin-zero states or can correspond to hyperglueballs, hyperonia, 
or pseudoscalar hypermesons.
For each of these cases, we make predictions for additional states that could be resonantly or pair produced and observed at the Large Hadron Collider or in future collider experiments. 
Heavy digauge boson resonances can provide a unified explanation for a number 
of small discrepancies and excesses in reported data from the Large Hadron Collider. 
\noindent}
\vspace{0.3cm}
\hrule

\vspace{0.6cm}

\end{titlepage}

\newpage 

\section{Introduction}

The search for new physics at the Large Hadron Collider (LHC) presents many challenges, including accurate theoretical predictions of Standard Model processes, data driven characterization of backgrounds, and modeling of detector responses.  All new physics searches exploit kinematic features to help distinguish signal from background. Perhaps the most effective of these is a relatively narrow resonance in the invariant mass spectrum of a fully reconstructable channel.  The discovery of the Higgs boson \cite{Aad:2012tfa, Chatrchyan:2012xdj} rested on such resonant digauge bosons channels.   In this paper we investigate some aspects of the physics that might underly additional digauge boson resonances many times heavier than the Higgs boson and that could be observed at the LHC. These considerations are particularly timely in light of recent reports from the ATLAS and CMS collaborations of a possible excess in the diphoton spectrum near 750 GeV \cite{CMS:2015dxe, ATLASdiphoton}. 

Among the possible resonant digauge boson channels, the diphoton channel is special in that the photon does not gain a mass from electroweak symmetry breaking, and at the renormalizable level couples only to electrically charged matter. Thus, obtaining a diphoton resonance at this level requires 
that charged matter either {\it constitutes} or {\it couples to} the resonant state. Couplings of the resonance to the known charged matter of the Standard Model could in principle play this role, 
but for a resonance many times heavier than the Higgs boson, direct two-body decay to these relatively light charged states would dominate over the diphoton decay mode and lead to a very small diphoton branching ratio. So, obtaining a diphoton resonance with significant strength several times heavier than the Higgs boson requires on general grounds additional charged matter that is either heavier than half the resonance mass or confined into bound states.  
If the additional matter is also charged under QCD, an analogous coupling of the resonance to digluons may be obtained. This provides a production mechanism through gluon fusion, as 
well as an associated resonance in the dijet channel. 

The general framework for a heavy $\gamma \gamma$
 resonance coming from additional charged and strongly interacting matter that either constitute or couple to the resonant state has important implications for other resonant final states that could be observed at the LHC.  
Primary among these are the closely related digauge boson final states $WW$, $ZZ$, 
$Z \gamma$ and $gg$. 
As we discuss below, the branching ratios of these 
digauge boson final states are related in a direct -- and for 
$Z \gamma$ and $gg$ in a rather robust -- way to the gauge representations of the additional matter.  
Observation of resonant final states coincident in mass with a diphoton signal 
could therefore provide rather model-independent information on the underlying 
matter representations. As we will see, whether a particular resonance provides evidence for all new matter representations or merely a subset depends on whether the resonance couples to, or is comprised of, the charged and colored matter. Furthermore, in the event that the resonance carries 
the quantum numbers of the vacuum, nothing forbids it from mixing with the 
Higgs boson, and such mixing will arise at some level irrespective of the underlying physics. 
As we show below, even a very small mixing with the Higgs can enhance the resonant branching ratios to the $WW$ and $ZZ$ final states, and also introduce $tt$ and $hh$ resonant final states. 

In this paper we formulate an experimental strategy for uncovering TeV-scale physics associated with a new resonance in a digauge boson channel. We first investigate some of the model-independent implications for resonant diboson (and possibly difermion) channels coming from electrically charged and strongly interacting matter associated with a resonance several times heavier than the Higgs boson. This provides guidelines for searching for other final states at the same invariant mass as the observed digauge boson resonance. The above-mentioned case of mixing with the Higgs provides new experimental opportunities but is also subject to experimental constraints from observed Higgs properties. We proceed to discuss constraints on, and observational consequences of, mixing between a new vacuum state and the Higgs boson. 

We then look away from the direct decay products of the heavy resonance to investigate possible additional signatures (resonant or not) associated with new degrees of freedom accompanying the resonance. To formulate predictions for signatures in nonresonant channels or resonant channels at other invariant masses, we construct a simple model framework that gives rise to heavy diphoton resonances in three distinct parametric limits: a weakly coupled limit in which the resonance is an elementary scalar or pseudoscalar whose SM couplings arise from loops of charged and colored matter, a strongly coupled limit in which the resonance is a glueball or quarkonium state of a confining gauge sector; and a strongly-coupled limit in which the resonance is a pion of a global symmetry spontaneously broken by strong dynamics.\footnote{The framework contains a variety of intermediate regimes and other limits that we will not fully explore here, including a quirky limit \cite{Kang:2008ea} with macroscopic strings and limits where the elementary scalar and new bound states mix appreciably.} Although the particular identity of the resonance varies in each regime, they share a range of common phenomena and point toward ancillary experimental signatures away from the diphoton invariant mass peak.

Taken together, searches for additional decay products of the heavy resonance, mixing effects with the Higgs boson, and additional states accompanying the resonance provide a concrete strategy for exploiting hints of new physics near the TeV scale. Our study of digauge boson resonances is quite general, but it does make a few assumptions that are consistent with the possible excess in the diphoton spectrum near 750 GeV, for example it is assumed that the diboson resonance corresponds to the lightest state in the spectrum of states beyond the SM. In what follows, where relevant we assume the diphoton events are genuine (as opposed to tetraphotons from the decay of light states). We use as our benchmark a resonance of mass 750 GeV with 
$\sigma \cdot {\rm Br}(pp \to \Phi \to \gamma \gamma)_{\rm LO}$ 
of 5 fb at $\sqrt{s} = 13$ TeV. As we will discuss further, for various reasons we do not attempt to accommodate the modest preference of ATLAS diphoton events for a large width.\footnote{A large number of recent preprints have studied the 750 GeV bump in a wide variety of contexts. Particularly relevant to our discussion are works analyzing phenomenological signals correlated with a diphoton excess~\cite{Mambrini:2015wyu,Buttazzo:2015txu,Franceschini:2015kwy,DiChiara:2015vdm,Cao:2015pto,Chao:2015ttq, Chakrabortty:2015hff,Falkowski:2015swt,Alves:2015jgx,Huang:2015evq,Berthier:2015vbb,Dey:2015bur,Dev:2015isx}, the general discussion of perturbative models in~\cite{shouldcitedraper, Gupta:2015zzs,Ahmed:2015uqt,Chao:2015nsm, Chang:2015bzc,Bardhan:2015hcr,Murphy:2015kag,deBlas:2015hlv,Chakraborty:2015gyj}, studies of quirks, quarkonia, and glueballs~\cite{Curtin:2015jcv,Agrawal:2015dbf}, and the considerable number of works on new pionlike states~\cite{Franceschini:2015kwy, Harigaya:2015ezk,Nakai:2015ptz,Molinaro:2015cwg,Matsuzaki:2015che,Bian:2015kjt,Bai:2015nbs,Cline:2015msi} (likewise in the context of composite Higgs models~\cite{Bellazzini:2015nxw,Low:2015qep,No:2015bsn,Belyaev:2015hgo}). Of course, the phenomenology related to this final state has been of interest even before the announcement of the excess near 750 GeV, see for example references \cite{Aguilar-Saavedra:2013qpa,Carpenter:2015gua}.}

This paper is organized as follows: In Section \ref{sec:independent} we formulate a variety of model-independent statements about resonant production and decay modes associated with a new heavy state with an appreciable diphoton branching ratio. This provides a set of guidelines for additional searches for final states with the same invariant mass. In Section \ref{sec:model} we then look away from the diphoton invariant mass and explore a range of additional signatures via a simple model framework whose various limits give rise to one or more heavy resonances. We offer some concluding remarks and comments on future directions in Section \ref{sec:conclusions}.

\section{Resonant production and decay modes} \label{sec:independent}

As discussed in the Introduction, an appreciable resonant signal in diphoton final states near the TeV scale is likely to involve matter charged under both electromagnetic and strong interactions, which either couples to the resonant state or constitutes the resonant state (e.g.~as a consequence of some confining dynamics). This leads to a variety of model-independent predictions for ancillary final states involving massless gauge bosons, principally $Z \gamma$ and $gg$. These predictions may be formulated most robustly in terms of ratios of these branching ratios relative to the $\gamma \gamma$ branching ratio. In the absence of mixing, similar predictions can be made for decays involving only massive gauge bosons, such as those to $WW$ and $ZZ$ final states. However, if the resonance carries the same quantum numbers as the vacuum, or if the resonance does not have well-defined parity and time-reversal quantum numbers, it may mix with the Higgs boson. This mixing principally alters branching ratios into massive gauge bosons $WW$ and $ZZ$ and introduces additional $tt$ and $hh$ resonant final states.

\begin{table}[h]
\begin{center}  
\begin{tabular}{cccccccc} 
     \multicolumn{2}{c}{Matter} &
& $ \tan \vartheta_{21}  $  
& $ \tan \vartheta_{13}   $  
&\\
   \multicolumn{2}{c}{Representation} 
& 
& 
 & \\ \\
$\bar{d}$ & $( \bar{\bf 3} , {\bf 1} )_{1 \over 3}$ 
    && 0  & 2/3  \\ \\
$\bar{u}$ & $( \bar{\bf 3} , {\bf 1} )_{- \! { 2 \over 3}}$ 
   &&  0  & 8/3  \\ \\
${\cal Q}$ & $( \bar{\bf 3} , {\bf 1} )_{1}$ 
    && 0  &  6  \\ \\
${Q}$ & $( {\bf 3} , {\bf 2} )_{1 \over 6}$ 
    &&  9  &  1/6    \\ \\
$\bar{d}, L$ & $( \bar{\bf 3} , {\bf 1} )_{1\over 3} \!  \oplus \! ({\bf 1} , {\bf 2})_{-\! {1\over 2}}$ 
   &&  3/5  & 5/3 \\ \\
$ Q , \bar{u} , $& $( {\bf 3} , {\bf 2} )_{1 \over 6} \! \oplus \! ( \bar{\bf 3} , {\bf 1} )_{-\! { 2 \over 3}} $ 
  &&   3/5  & 5/3    \\
 $ \ell $ & $ ( {\bf 1} , { \bf 1})_1$ & \\ \\
$ Q , \bar{u} , $ & $~ \! ( {\bf 3} , {\bf 2} )_{1 \over 6} \! \oplus \! ( \bar{\bf 3} , {\bf 1} )_{-\! { 2 \over 3}} $ 
  &&  3/5  & 5/3   \\
$\bar{d}, L, $ & $  ( \bar{\bf 3} , {\bf 1} )_{1\over 3} \! \oplus \!  ({\bf 1} , {\bf 2})_{-\! {1\over 2}}$ & \\
 $ \ell$  & $ ( {\bf 1} , { \bf 1})_1$ & \\ \\ 
 \end{tabular}
\caption{Weak-hypercharge and hypercharge-strong 
gauge interaction weighted 
amplitude ratios for either the scalar or pseudoscalar 
couplings of a single spin-zero state to two gauge boson states
originating from various
$SU(3)_C \times SU(2)_L \times U(1)_Y$ 
matter representations 
in the absence of electroweak symmetry breaking or form factor effects. All $\kappa_{\alpha}$ have been taken to be one.
For multiple representations these ratios are obtained with 
uniform coupling of the resonance to each representation
in either the degeneracy or decoupling 
limits of the massive matter spectrum.
In these limits any 
combination of matter representations that can be embedded within complete multiplets of a non-Abelian grand unified gauge group have $\tan \vartheta_{21}  = 3/5$ and 
$ \tan \vartheta_{13} = 5/3$. }
\label{tab:angles} 
\end{center}
\end{table}

In this section we systematically study the general, model-independent properties of a neutral 
spin-zero resonance decaying to digauge boson final states. In the absence of mixing with the Higgs boson, the production cross section and all ratios of branching ratios can be parametrized in terms of angles $\vartheta_{ij}$ which 
for charged states with equal spins and a degenerate spectrum are given by
\begin{eqnarray}
\tan \vartheta_{21} &=& { {\rm Tr} \big(  
    \, C(R_2)   
         \,   \big)  \over 
    {\rm Tr} \big( \, Y^2
                         \, \big) }\label{eq:theta21}
        \\
\tan \vartheta_{13} &=& 
{ {\rm Tr}  \big(  \, Y^2        \,   \big)  \over 
    {\rm Tr} \big( \, C(R_3) 
                \, \big) }
                \label{eq:theta13}
\end{eqnarray}
where the trace is over Standard Model representations 
$R_3 \otimes R_2$ of 
$SU(3)_C \times SU(2)_L$, $U(1)_Y$ hypercharge is normalized as $Y=Q - T_3$
and $C(R_{2,3})$ denotes the index of representations of $SU(2)_L$ and $SU(3)_C$ defined as ${\rm Tr}(T^a T^b) = C(R) \delta^{ab}$
with the normalization $C(\Box) = {1 \over 2}$.
Examples of $\vartheta_{ij}$ for some common choices of representations are given in Table~\ref{tab:angles}. Note that for complete multiplets of a non-Abelian grand unified gauge group if the resonance has equal couplings to each representation (the last three entries in the table), the $\vartheta_{ij}$ take on universal values.

\begin{table}[h]
\begin{center}  
\begin{tabular}{cccccccc} 
     \multicolumn{2}{c}{Matter} 
& $ \underline{{\rm Br}( \phi \! \to \! Z \gamma)} \! $  
& $ \underline{{\rm Br}( \phi \! \to \! WW)} \! $  
& $ \underline{{\rm Br}( \phi \! \to \! ZZ)} \! $  
& $ \underline{{\rm Br}( \phi \! \to \! \gamma \gamma)} \! $  
&\\
   \multicolumn{2}{c}{Representation} 
& ${\rm Br}( \phi \! \to \! \gamma \gamma) \! $
& ${\rm Br}( \phi \! \to \! \gamma \gamma) \! $
& ${\rm Br}( \phi \! \to \! \gamma \gamma) \! $
& ${\rm Br}( \phi \! \to \! gg) \! $
 & \\ \\
$\bar{d}$ & $( \bar{\bf 3} , {\bf 1} )_{1 \over 3}$ 
   &  0.61 & 0  & 0.093 &  $ 4.1 \times 10^{-4}$  \\ \\
$\bar{u}$ & $( \bar{\bf 3} , {\bf 1} )_{- \! { 2 \over 3}}$ 
   & 0.61   & 0  & 0.093 &  $ 6.6  \times 10^{-3}$   \\ \\
${\cal Q}$ & $( \bar{\bf 3} , {\bf 1} )_{1}$ 
   & 0.61   & 0  & 0.093 &  $ 3.3 \times 10^{-2}$   \\ \\
${Q}$ & $( {\bf 3} , {\bf 2} )_{1 \over 6}$ 
   & 4.5   & 26  & 7.4 &  $ 2.6 \times 10^{-3}$   \\ \\
  \multicolumn{2}{c}{Unified} ~~~
   & 0.18  & 4.5  & 1.7 &  $ 6.6 \times 10^{-3}$   \\ \\
 \end{tabular}
\caption{Leading-order ratio of branching ratios
of digauge boson final states 
for a spin-zero resonance $\phi$ with gauge boson interactions originating 
from various
$SU(3)_C \times SU(2)_L \times U(1)_Y$ 
matter representations
in the absence of electroweak symmetry-breaking effects or 
mixing with the Higgs boson. 
Unified refers to any combination of matter representations that can be 
embedded within complete multiplets of a non-Abelian grand unified gauge group. 
With a single matter representation 
these ratios of branching ratios are independent of the massive matter spectrum, 
and with multiple representations are obtained 
with uniform coupling of the resonance 
to each representation in either the degeneracy or decoupling 
limits of the massive matter spectrum. 
The branching ratios ${\rm Br}( \phi  \to  W W)$ and 
${\rm Br}( \phi  \to  Z Z)$ are very sensitive to mixing effects with the Higgs boson. 
The gauge couplings are evaluated at a renormalization scale of $\mu = 750$ GeV. 
}
\label{tab:brs} 
\end{center}
\end{table}

To see how these ratios enter into physical predictions, consider first the decays of the resonance to final states that include one or more massless gauge bosons. To leading order, these decays are independent of mixing with the electroweak symmetry-breaking sector.  In this respect they provide a rather direct probe of the quantum numbers of the charged matter connecting the resonance to SM gauge bosons. Although absolute branching ratios cannot be reliably determined, the ratios of branching ratios into these final states can be uniquely represented in terms of the matter representations of the additional charged and 
strongly interacting
 states. In particular, the ratio of branching ratios into $Z \gamma$ and $\gamma \gamma$  is given by
\begin{eqnarray}
{ {\rm Br}(\phi \to Z \gamma ) \over {\rm Br}(\phi \to \gamma \gamma ) } = 
 2 \, \bigg( { \tan \theta_W - \tan \vartheta_{21} \cot \theta_W \over 
  1 + \tan \vartheta_{21}  }   \bigg)^2 
  \bigg( 1 - {m_Z^2 \over m_\phi^2 } \bigg)^3
\end{eqnarray}
Note that this depends on only a single ratio of amplitude coefficients, namely $\tan \vartheta_{21}$, so that a measurement of the ratio of branching ratios to $Z \gamma$ and $\gamma \gamma$ would fully determine this parameter. Examples of the numerical values of this and other ratios for various canonical matter representations are given in Table \ref{tab:brs}.
Likewise, the ratio of branching ratios into $\gamma \gamma$ and $gg$ is given by
\begin{eqnarray}
{ {\rm Br}(\phi \to \gamma  \gamma ) \over {\rm Br}(\phi \to gg ) } = 
 { \alpha^2 \over 8 \, \alpha_s^2} 
  \tan^2 \vartheta_{13} \, \big( 1 + \tan \vartheta_{21} \big)^2
\end{eqnarray}
and depends on two ratios of amplitude coefficients, $\tan \vartheta_{21}$ and $\tan \vartheta_{13}$. In principle, given a measurement of $Z \gamma, \gamma \gamma,$ and $gg$ branching ratios, both $\tan \vartheta_{13}$ and $\tan \vartheta_{21}$ could be unambiguously determined. 
For a pure scalar state with interactions that are even under time reversal, the ratio of branching ratios 
into $WW$ and $\gamma \gamma$
in the absence of mixing with the Higgs boson are 
\begin{equation}
 {  {\rm Br}(\phi \to WW ) \over {\rm Br}(\phi \to \gamma \gamma ) } = 
 { 2 \, \tan^2 \vartheta_{21}  \over 
  \sin^4 \theta_W \, (1 + \tan \vartheta_{21})^2  }   
  \bigg( 1 - { 4 m_W^2 \over m_\phi^2 } + { 6 m_W^4 \over m_\phi^4} \bigg)
  \bigg(1 -  { 4 m_W^2 \over m_\phi^2 } \bigg)^{1/2}
\end{equation}
and likewise $ZZ$ and $\gamma \gamma$ 
\begin{equation}
 {  {\rm Br}(\phi \to ZZ ) \over {\rm Br}(\phi \to \gamma \gamma ) } = 
 \bigg( { \tan^2 \theta_W + \tan \vartheta_{21} \cot^2 \theta_W \over 
  1 + \tan \vartheta_{21}  }   \bigg)^2 
    \bigg( 1 - { 4 m_Z^2 \over m_\phi^2 } + { 6 m_Z^4 \over m_\phi^4} \bigg)
  \bigg(1 -  { 4 m_Z^2 \over m_\phi^2 } \bigg)^{1/2}
\end{equation}
The ratios for pure pseudoscalar differ only in the numerically unimportant phase 
space factors. 
These ratios are also determined only by $\tan \vartheta_{21}$ in the absence of mixing 
with the Higgs.  
So a measurement of these along with branching 
ratios to $Z \gamma$ and $\gamma \gamma$
provides a very robust test for any such mixing.

In the narrow width approximation, the partial width to gluons may also be related to the production cross section via
\begin{eqnarray}
\sigma(pp \to \phi) = { \pi^2 \over 8 \,  s_{pp}} \,  {\Gamma( \phi \to gg) \over m_\phi} 
\int_{\tau_0}^1 {d x \over x} ~ f_g(x) f_g(\tau_0 / x) 
\end{eqnarray}
where $\tau_0 = m_\phi^2 / s_{pp} $ and $f_g(x)$ is the gluon 
parton distribution function. 
This provides an additional handle on the couplings of the resonance to massless Standard Model gauge bosons beyond the ratios of branching ratios.  Numerical values of the resonant production cross section at $\sqrt{s} = 13$ TeV for various canonical matter representations are given in Table \ref{tab:channels}.

\begin{table}[h]
\begin{center}  
\begin{tabular}{cccccccc} 
     \multicolumn{2}{c}{Matter} 
& $ \underline{ \Gamma( \phi \! \to \! gg)_{\rm LO}  /  {\rm GeV} } $  
& $ \underline{ ~\sigma \! \cdot \! {\rm Br}( pp \! \to \! \phi \! \to \! gg)_{\rm LO}  /  {\rm pb} ~} $  
&\\
   \multicolumn{2}{c}{Representation} 
& $  \sigma \! \cdot \! {\rm Br}( pp \! \to \! \phi \! \to \! \gamma \gamma)_{\rm LO} / 5 \, { \rm fb} $
& $  \sigma \! \cdot \! {\rm Br}( pp \! \to \! \phi \! \to \! \gamma \gamma)_{\rm LO} / 5 \, { \rm fb} $ 
   & \\ \\
$\bar{d}$ & $( \bar{\bf 3} , {\bf 1} )_{1 \over 3}$ 
  & 1.8 & 12 \\ \\
$\bar{u}$ & $( \bar{\bf 3} , {\bf 1} )_{- \! { 2 \over 3}}$ 
  & 0.12  & 0.76 \\ \\
${\cal Q}$ & $( \bar{\bf 3} , {\bf 1} )_{1}$ 
  & 0.023 & 0.15  \\ \\
${Q}$ & $( {\bf 3} , {\bf 2} )_{1 \over 6}$ 
  & 0.30 & 1.9  \\ \\
  \multicolumn{2}{c}{Unified} ~~~
  & 0.12  & 0.76 \\ \\
 \end{tabular}
\caption{Leading-order resonant partial decay width to gluons and 
resonant digluon cross section for the benchmark 
spin-zero resonance with mass $m_\phi = 750$ GeV and 
with gauge boson interactions originating from various matter representations
and with diphoton cross section 
$\sigma \cdot {\rm Br}(pp \! \to \phi \to \gamma \gamma)_{\rm LO} = 5$ fb 
at $\sqrt{s} = 13$ TeV using ratio of branching ratios 
given in  Table \ref{tab:brs}.  
The total resonant decay width is 
$\Gamma( \phi \to {\rm All}) = \Gamma( \phi  \to  gg) /  {\rm Br}( \phi  \to  gg)$.
}
\label{tab:channels} 
\end{center}
\end{table}

Although the production rate and ratios of branching ratios of a resonance provide a sensitive probe of the charged matter connecting it to the Standard Model, whether or not the existence of {\it all} new charged and/or strongly interacting matter can be inferred from the production and decays of resonances depends on whether the resonances are comprised of, or merely couple to, the new matter. If resonances are comprised of the matter, as in the case of bound states, the only resonances observable at the LHC will be those made at least in part from strongly interacting constituents. There may be additional resonances comprised of charged matter
without strong QCD interactions, 
but these are likely to be produced too weakly at the LHC to be observable. In contrast, if resonances merely couple to the new matter, then all the states interacting with the resonance leave their imprint on its production rate and branching ratios. This dichotomy between matter constituting a resonance or coupling to a resonance will be borne out clearly by explicit examples in the following section.

Whatever the precise nature of the resonance, several features of its decays to $WW$ and $ZZ$ final states warrant further consideration.  In the absence of mixing with the Higgs, the ratios of branching ratios involving massive gauge bosons are determined in terms of the $\vartheta_{ij}$ angles much as in the massless case. As is apparent from Table~\ref{tab:brs}, if the new matter fields are charged under $SU(2)_L$, the branching ratio to $WW$ can be significantly enhanced relative to $\gamma \gamma$. For example, a 750 GeV diphoton resonance induced by matter transforming in unified representations with a 13 TeV rate of $\sigma \cdot {\rm Br}(pp \to \phi \to \gamma \gamma)_{\rm LO} \simeq 5$ fb would have a corresponding rate for resonant $WW$ production of $\sigma \cdot {\rm Br}(pp \to \phi \to WW)_{\rm LO} \simeq 22.5$ fb at $\sqrt{s} = 13$, 
or $\sigma \cdot {\rm Br}(pp \to \phi \to WW)_{\rm LO} \simeq 5$ fb at $\sqrt{s} = 8$ TeV. 
This is 
interesting in light of a suggestive excess in semileptonic $WW$ events at 750 GeV in 8 TeV CMS data \cite{Khachatryan:2015cwa}, and suggests that $WW$ is a particularly promising channel for ongoing searches at 13 TeV.

These conclusions are substantially altered when the resonance mixes with the Higgs to even a modest degree. In the absence of mixing, $WW$ and $ZZ$ final states involve contributions from both longitudinally and transversely polarized gauge bosons, but the contribution of longitudinal polarizations to the $WW$ and $ZZ$ partial widths is suppressed by $\mathcal{O}(m_Z^4/m_\phi^4)$ relative to the transverse polarizations. Mixing with the Higgs, however, introduces tree-level couplings to  longitudinal components of $W$ and $Z$ bosons that can significantly modify branching ratios to $WW$ and $ZZ$ even when the mixing angle is small. The resulting mixing-induced partial widths for the decay of the resonance to longitudinal $W$ and $Z$ bosons scale as $\Gamma(\phi \to A_L A_L) \propto g^2 \sin^2 \theta_{\phi h} ~{m_\phi^3}/{m_A^2}$, where $\theta_{\phi h}$ is the mixing angle between the 750~GeV resonance and the physical Higgs boson. For mixing angles of natural size $\sin \theta_{\phi h} \sim m_h / m_\phi$, the mixing angle suppression is largely compensated for by enhanced longitudinal couplings. Although Higgs mixing is somewhat constrained by Higgs coupling measurements, even small resonance-Higgs mixing can significantly enhance the longitudinal decay modes of the resonance and alter the branching ratios to $WW$ and $ZZ$ relative to those involving one or more massless gauge bosons. This makes ratios of branching ratios to $WW$ and $ZZ$ final states -- and measurement of the longitudinal versus transverse polarization fractions -- a particularly sensitive probe of mixing between the Higgs and a new resonance.

In addition to enhancing decays into $WW$ and $ZZ$ final states, mixing with the Higgs also introduces di-Higgs and ditop resonant channels. While these may be appreciable, they are not enhanced by longitudinal couplings, and so resonant $tt$ and $hh$ production is likely to provide less sensitivity than $WW$ and $ZZ$. Nonetheless, this motivates continued searches for resonant di-Higgs production, with resonant $tt$ production being far more challenging on account of the large Standard Model background.

Of course, it is entirely possible that the resonant state has a conserved quantum number that distinguishes it from the vacuum and forbids mixing with the Higgs.  For example, if parity and time reversal are approximate symmetries in the resonant sector and the resonance is a pseudoscalar that is odd under these symmetries, then mixing with the Higgs is suppressed.\footnote{The parity properties of the resonance can also be extracted from $WW$ and $ZZ$ decays, including time-reversal violating 
interference effects between the two 
using kinematic correlations within the fully reconstructed final states.} If such mixing with the Higgs is suppressed because of the structure of physics underlying the resonance, then the ratio of branching ratios to $WW$ and $ZZ$ final states are determined in terms of the same ratio of amplitude coefficients that appears in the expression for $Z \gamma$.  So in the absence of significant form factor effects, there is a one-family parameter of predictions relating these three ratios of branching ratios in this case. This overdetermined set of observables provides a very good test for the absence of mixing of the resonance with the Higgs boson.  

Even in the event that mixing between the Higgs and a diphoton resonance is forbidden or suppressed, there could be other states within the same sector that {\it do} mix substantially with the Higgs and have enhanced $W_L W_L$ and $Z_L Z_L$ resonant final states. (As we will see, this is often borne out in realistic models explored in Section \ref{sec:model}.) If these states are even heavier than the diphoton resonance, then the increase in their production cross sections between Run 1 and Run 2 of the LHC can be substantial. For example, for resonances of, say, 1.8 TeV and 2.6 TeV the cross section ratios are $\sigma_{13 \, {\rm TeV}}/\sigma_{8 \, {\rm TeV}} \sim 11$ and $\sigma_{13 \, {\rm TeV}}/\sigma_{8 \, {\rm TeV}} \sim 24$, respectively.  If these states had gluon couplings comparable to those of the potential diphoton resonance (i.e. $\Gamma(\phi \to gg) / m_\phi \sim 3 \times 10^{-4}$), their 13 TeV total cross sections would respectively be on the order of 12 fb and 1 fb.

\section{The Physics underlying scalar resonances} \label{sec:model}

The general model-independent treatment of the previous section is suitable for characterizing additional possible decay modes of a heavy resonance. However, as has already been observed, the properties of the resonance require additional degrees of freedom to provide a portal between the scalar and Standard Model gauge bosons. These additional degrees of freedom give rise to new experimental signatures that may be pursued in different channels and at different mass scales. Detailed understanding of these new signatures requires a specific model framework.

In this section we present a model framework consisting of a minimal set of ingredients giving rise to suitable high-mass diphoton resonances in various regions of parameter space. These ingredients consist of an elementary scalar or pseudoscalar $\Phi$; new portal matter (either fermions or scalars) charged under the Standard Model; and possibly new gauge bosons that gauge an additional global symmetry of the portal states. We will refer to the new gauge group as ``hypercolor'' and indicate quantities in the hypercolor sector with a subscript $HC$.

There are three relevant regimes that are explored by varying the hypercolor coupling $g_{HC}$ (equivalently, the confinement scale $\Lambda_{HC}$ associated with $g_{HC}$), the mass scale $M$ of the portal matter, and the mass scale $M_\Phi$ of the elementary (pseudo)scalar:
\begin{enumerate}
\item In the limit $g_{HC} \to 0$ ($\Lambda_{HC} \to 0$), $M_\Phi \simeq 750$ GeV, and $M > M_\Phi / 2$, hypercolor becomes purely a global symmetry and the observed diphoton resonance corresponds to the production and decay of the elementary $\Phi$ state. Here the resonance merely couples to the new charged and colored matter; production and decay modes of the resonance arise from loops of portal states.
\item In the limit $M_\Phi \to \infty$ and $\Lambda_{HC} \sim M$, the elementary (pseudo)scalar can be entirely eliminated and the diphoton resonance arises as a bound state of strong hypercolor dynamics. The resonance may be a hyperglueball (i.e. a bound state of hypergluons) with a mass determined largely by $\Lambda_{HC}$; a hyperonium state (a bound state of portal fields) with mass determined largely by $M$; or an admixture of the two. If the resonance is a hyperglueball, it couples to the new charged and colored matter, and its production and decay rates reflect the quantum numbers of the portal states. On the other hand, if the resonance is a hyperonium state, it is instead comprised of the charged and colored matter; production and decay rates reflect only the matter states forming the resonance.
\item In the limit $M_{\Phi}  \to \infty$ and $M \ll \Lambda_{HC}$, the elementary (pseudo)scalar can again be eliminated. If hypercolor confinement spontaneously breaks a global symmetry, the diphoton resonance arises as a hyperpion whose mass is determined by the geometric mean of $M$ and $\Lambda_{HC}$. As in the case of hyperonia, resonances are comprised of the charged and colored matter.
\end{enumerate}

\begin{figure}[h] 
   \centering
   \includegraphics[width=3in]{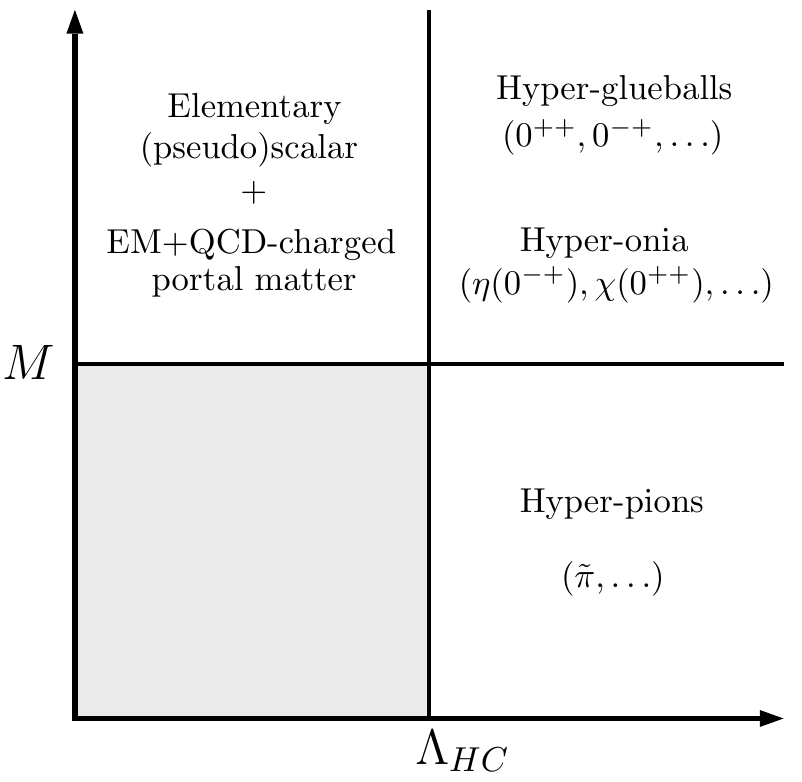} 
   \caption{Relevant regimes of the model framework as a function of the portal matter scale $M$ and the hypercolor scale $\Lambda_{HC}$. We profile over the mass of the elementary (pseudo)scalar $\Phi$.}
   \label{fig:regimes}
\end{figure}

These regimes are illustrated in Figure \ref{fig:regimes}. In each regime there is a variety of additional signatures associated with the relevant degrees of freedom, which lead to a variety of promising experimental strategies. Of course, there are other mixed regimes of this framework that may also provide interesting signals; for the sake of clarity we focus on the above three limits in this work.

\subsection{Elementary scalars}

Consider first the weakly coupled limit, in which the diboson resonance is an elementary scalar or pseudoscalar that is neutral under the Standard Model. The (pseudo)scalar couples to electromagnetically charged and strongly interacting portal matter, which may be fermions and/or scalars. The portal states may acquire some or all of their masses from an expectation value of the scalar, although this is not necessary for them to efficiently connect the resonance to Standard Model bosons. We assume all these portal states are heavier than half the resonance mass, $M > M_\Phi / 2$, so that the resonance cannot decay on-shell into pairs of portal states. The hypercolor gauge symmetry merely becomes a flavor symmetry of the portal states in the limit in the weakly coupled limit $g_{HC} \to 0$, so we allow for $N_f$ flavors of portal states transforming in a given Standard Model representation. 

Loops of portal states then connect the (pseudo)scalar resonance to Standard Model bosons, providing an avenue for production and decay of the resonance at the LHC. In the case of a scalar $\Phi$ we may consider couplings to (Dirac) fermionic and (complex) scalar portal matter of the form
\begin{eqnarray}
\mathcal{L}_{0^{++}} \supset y_\Psi \Phi \bar \Psi \Psi + y_\varphi \Phi |\varphi|^2
\end{eqnarray}
where $\Psi, \overline \Psi$ are vectorlike fermions and $\varphi$ are complex scalars. 
These couplings lead to partial widths for the decay of the scalar into Standard Model bosons. In particular, the partial widths for the decay of the scalar into pairs of gluons or photons are
\begin{eqnarray}
\Gamma(\Phi \to gg) &
 \! \! \! = \! \! \! 
    & \frac{\alpha_s^2}{32 \pi^3} \frac{M_\Phi^3}{M^2}  \left| \sum_i 
       \left[ \frac{4}{3} N_f  
 C(R_3)
y_\Psi f_{1/2} \! \left( \frac{4 M^2}{M_\Phi^2} \right) +  \frac{1}{6} N_f 
    C(R_3)
 \frac{y_\varphi}{M} f_0 \! \left( \frac{4 M^2}{M_\Phi^2} \right) \right] \right|^2 \\
\Gamma(\Phi \to \gamma \gamma) 
& \! \! \! = \! \! \! & 
\frac{\alpha^2}{256 \pi^3} \frac{M_\Phi^3}{M^2} \left| \sum_i \left[ \frac{4}{3} N_f d_R Q_i^2  y_\Psi f_{1/2} \! \left( \frac{4 M^2}{M_\Phi^2} \right) +  \frac{1}{6} N_f d_R Q_i^2 \frac{y_\varphi}{M} f_0 \! \left( \frac{4 M^2}{M_\Phi^2} \right) \right] \right|^2
\end{eqnarray}
where the sums run over all portal fermions and scalars, $d_R$ is the dimension of the $SU(3)$ representation of a given portal state, $C(R_3)$ is the index of the corresponding $SU(3)$ representation as defined in equation~\ref{eq:theta13}, and the loop functions $f_{1/2}, f_0$ are \cite{Gunion:1989we}
\begin{eqnarray}
f_{1/2}(\tau) &=& \frac{3 \tau}{2} \left[ 1 + (1-\tau) [\sin^{-1}(1/\sqrt{\tau})]^2 \right] \\
f_{0}(\tau) &=& -3 \tau \left[ 1 - \tau [\sin^{-1}(1/\sqrt{\tau})]^2 \right]
\end{eqnarray}
which are normalized such that they asymptote to $f_i \to 1$ in the limit $2 M \gg M_\Phi$. Ratios of these partial widths bear out the model-independent features discussed in Section \ref{sec:independent}, with the loop functions playing the role of form factors.

In the case of a pseudoscalar resonance we consider couplings to fermionic portal matter of the form
\begin{eqnarray}
\mathcal{L}_{0^{-+}} \supset y_{\Psi} \Phi \bar \Psi i \gamma_5 \Psi 
\end{eqnarray}
These couplings lead to partial widths for the decay of the pseudoscalar into pairs of gluons and photons of the form
\begin{eqnarray}
\Gamma(\Phi \to gg) &=& \frac{\alpha_s^2}{32 \pi^3} \frac{M_\Phi^3}{M^2}  \left| \sum_i 2 N_f C(R_3) y_\Psi g_{1/2} \! \left( \frac{4 M^2}{M_\Phi^2} \right) \right|^2 \\
\Gamma(\Phi \to \gamma \gamma) &=&  \frac{\alpha^2}{256 \pi^3} \frac{M_\Phi^3}{M^2} \left| \sum_i  2 N_f d_R Q_i^2 y_\Psi g_{1/2} \! \left( \frac{4 M^2}{M_\Phi^2} \right) \right|^2
\end{eqnarray} 
where in this case the loop function 
\begin{eqnarray}
g_{1/2}(\tau) &=& \tau [\sin^{-1}(1/\sqrt{\tau})]^2
\end{eqnarray}
is normalized such that $g_{1/2} \to 1$ for $2 M \gg M_\Phi$. Again, the ratio of these partial widths (and additional partial widths into other Standard Model gauge bosons)  exemplifies the features discussed in Section \ref{sec:independent}.

\begin{figure}[t] 
   \centering
   \includegraphics[height=1.9in]{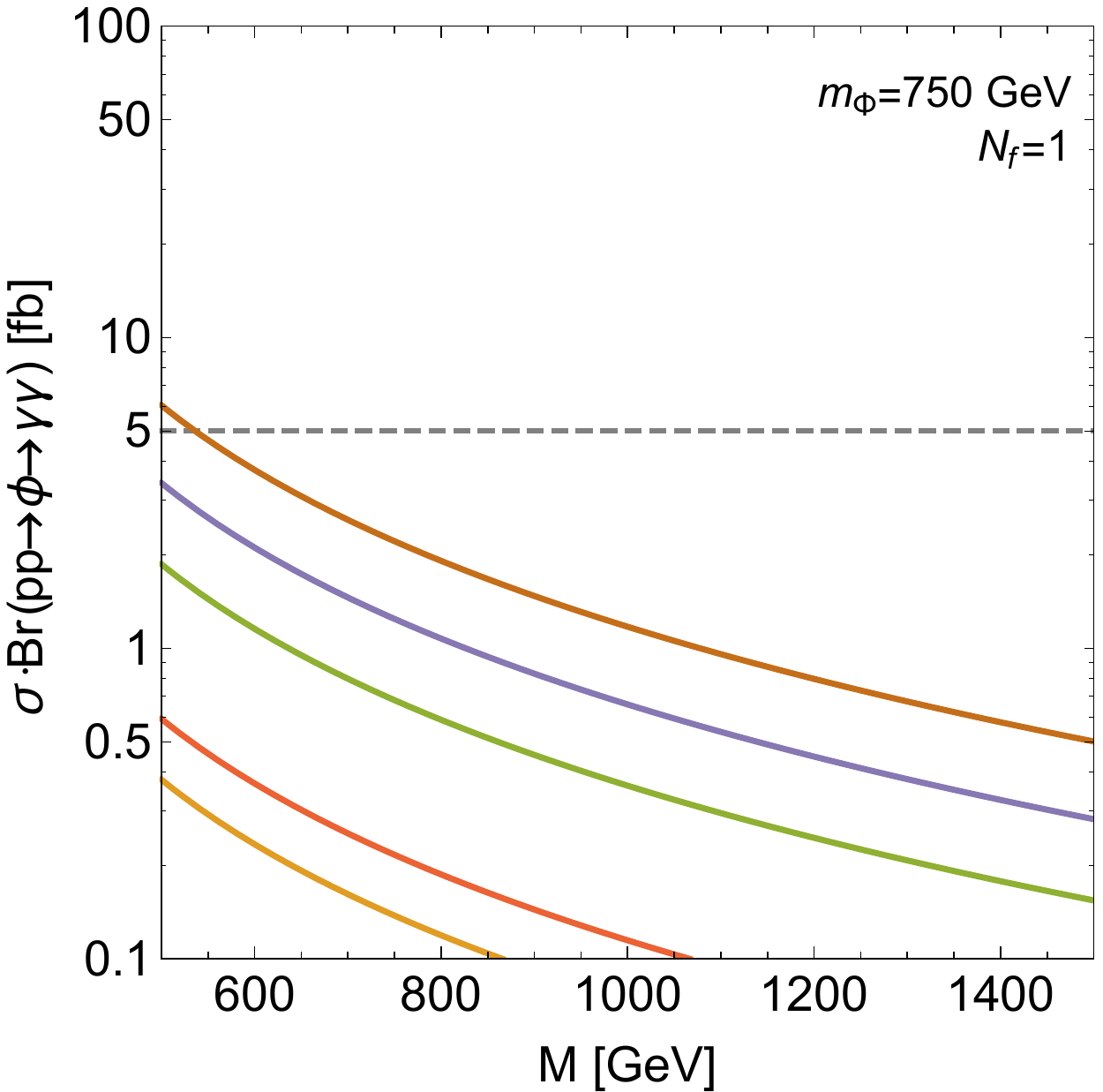} 
    \includegraphics[height=1.9in]{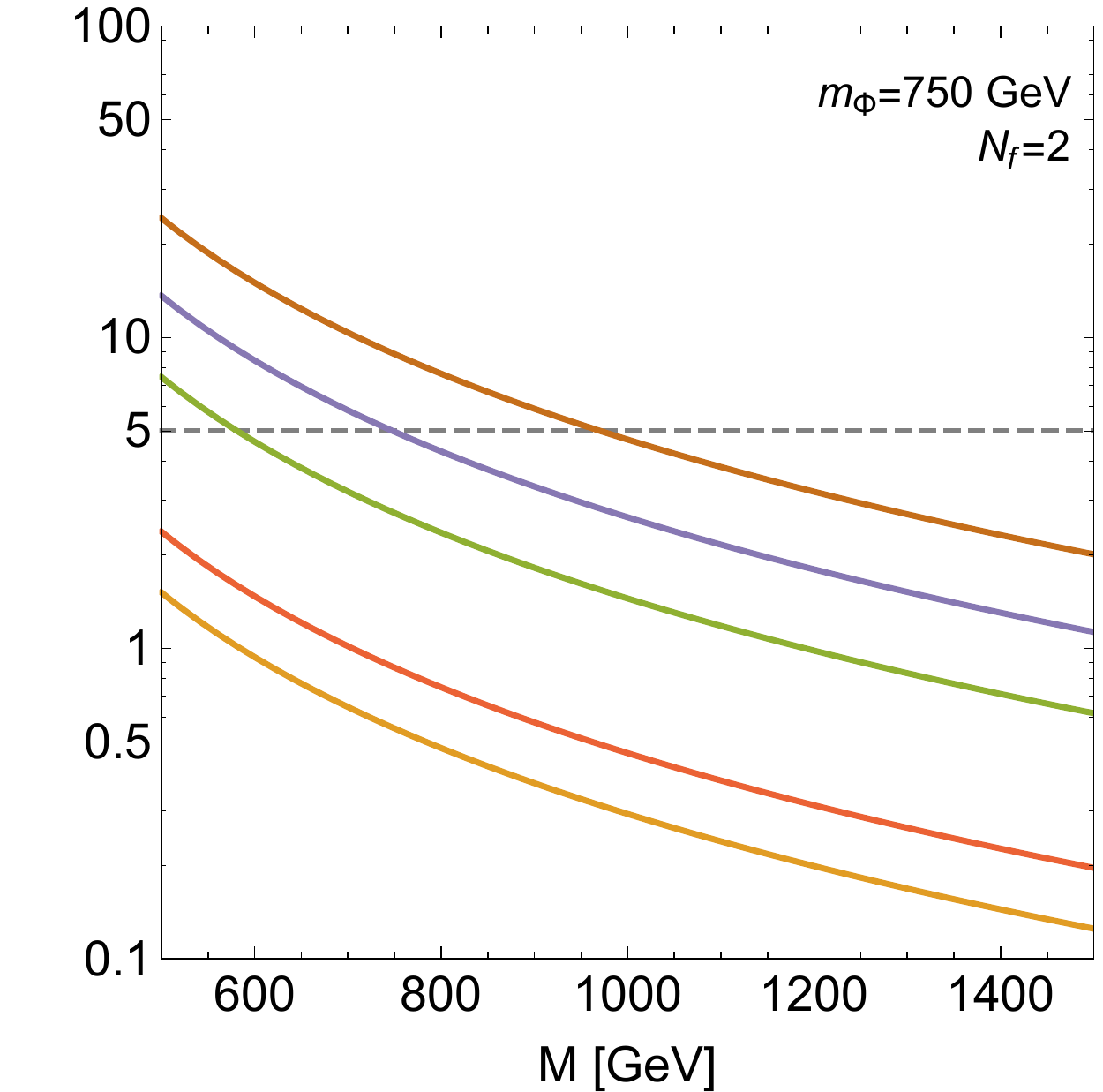} 
     \includegraphics[height=1.9in]{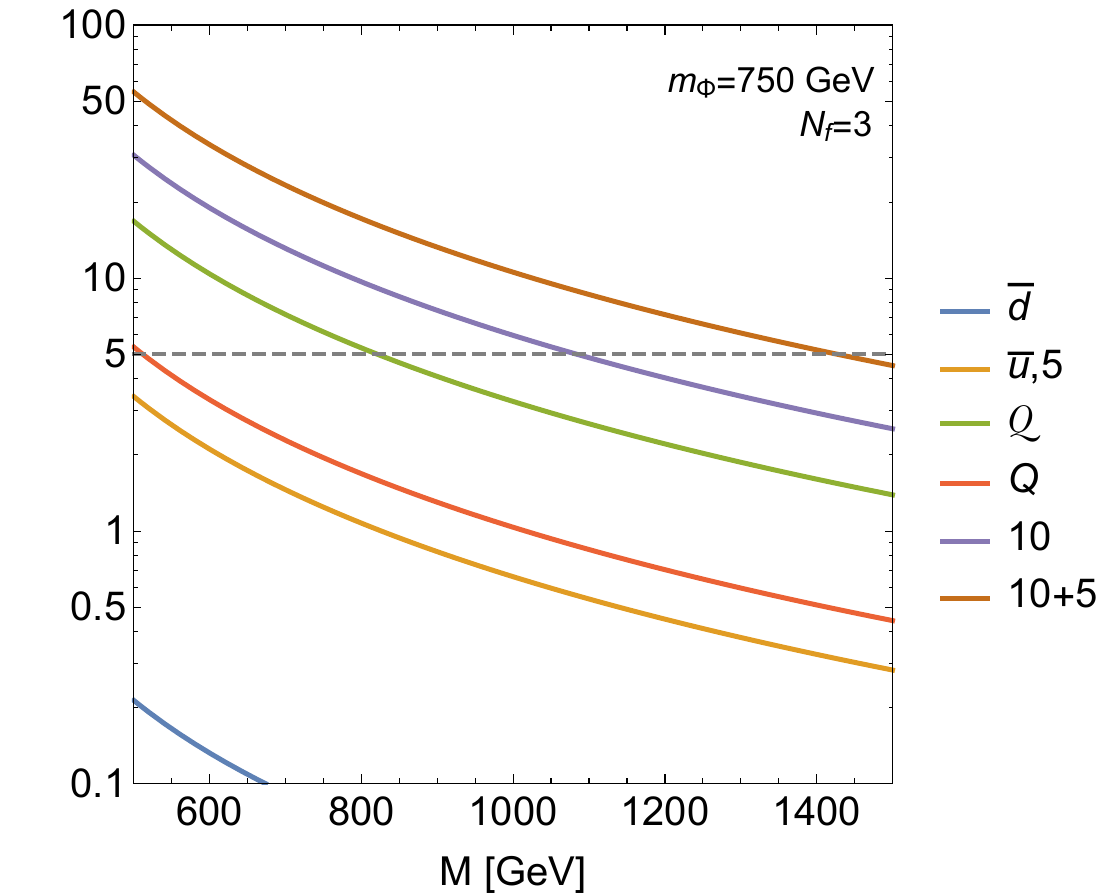} 
   \caption{The leading-order proton-proton 
   cross section times diphoton branching ratio in fb at $\sqrt{s} = 13$ TeV for 
   an elementary spin-zero scalar resonance of mass 750 GeV coupled at one-loop to the Standard Model 
   gauge bosons through various vectorlike representations of fermionic portal matter fields of mass $M$. 
   Mixing with the Higgs boson has been set to zero, and the Yukawa couplings have been set to $y_\Psi=1$. 
      Contours are shown for vectorlike fermions $\bar d;$ $\bar u;$ $\mathcal{Q};$ $Q;$ ${\bf 5} \equiv \bar d + L;$ ${\bf 10} \equiv Q +\bar u + \ell;$ and ${\bf 10} + {\bf 5} \equiv Q + \bar u + \bar d + L + \ell$ defined in Table \ref{tab:angles}. The dashed gray line denotes $\sigma \cdot {\rm Br}(pp \to \Phi \to \gamma \gamma) = 5$ fb.}
   \label{fig:sigbrgaga}
\end{figure}

In general, the observation of a resonant signal at the LHC would suggest a preferred mass scale and couplings for portal states. For example, the 13 TeV resonant production cross section of a scalar whose gluon couplings are induced by $N_f$ vectorlike fermionic flavors transforming in the fundamental of QCD is
\begin{eqnarray} \label{eq:xsec}
\sigma(pp \to \phi) \simeq 10 \, {\rm fb} \times N_f^2 y_\Psi^2 \left( \frac{1 \, {\rm TeV}}{M} \right)^2
\end{eqnarray}
Note that the cross section due to scalar portal fields is nearly an order of magnitude smaller.

A diphoton excess near 750 GeV would generically favor portal states at or beneath the TeV scale, with the precise scale and couplings depending on the full set of quantum numbers of portal states. A broad variety of cases can be worked out (for a single copy of given matter representations) simply using (\ref{eq:xsec}) in conjunction with the results in Table \ref{tab:brs}. To illustrate the range of parameter space, Figure \ref{fig:sigbrgaga} shows the cross section times branching ratio to diphotons for a 750 GeV scalar resonance with $N_{f} = 1,2,3$ copies of fermionic portal matter in various Standard Model representations.  \\

\noindent {\bf Additional Signatures:} \\

A diboson resonance visible at the LHC constrains the masses of the portal states to lie near the TeV scale. This provides strong signal-based motivation to search for Standard Model-charged states near the weak scale (in contrast with the theory-based motivation of naturalness arguments). The signatures of portal states vary depending on their quantum numbers and possible additional couplings to the Standard Model. Colored portal states will be pair produced through QCD processes, while electroweak portal states will be pair produced via Drell-Yan. The decays depend sensitively on the quantum numbers of the state. For vectorlike portal fermions, the simplest possibility is for the fermions to mix with Standard Model fermions through modest Yukawa couplings. This provides an avenue for the fermions to decay into Standard Model final states with the full range of signatures expected of vectorlike quarks. When quantum numbers permit, these states may also couple to a Standard Model fermion plus a stable neutral scalar, leading to missing energy signatures. Alternately, if quantum numbers forbid tree-level mixing with Standard Model fermions, these states may decay via higher-dimensional operators. For portal scalars, the decay modes also depend on their quantum numbers. They may couple to pairs of Standard Model fermions or to a Standard Model fermion plus a light neutral fermion, leading to missing energy in scalar decays. Signatures of these states cover the range of signals in $R$-parity-violating and $R$-parity-conserving  supersymmetry, respectively. Although there are many exotic possibilities for the decays of these states, in general the signature space is covered by (and strongly motivates) ongoing searches for scalar and fermionic partner states -- with various Standard Model quantum numbers -- at the LHC.

\subsection{Hyperglueballs and Hyperonia}

Moving away from the weakly coupled limit, it is possible to entirely eliminate the elementary (pseudo)scalar and instead focus on phenomena associated with a strong hypercolor gauge sector.  Strong hypercolor interactions can give rise to a variety of suitable bound states. Bound states of hypergluons may lie at the bottom of the hyperspectrum, with $J^{PC} = 0^{++}, 0^{-+},$ or even $2^{++}$ glueballs giving rise to appreciable diphoton signals.\footnote{\label{fn:parity}This classification of states corresponds to the special slice of parameter space where parity is conserved. {\it A priori}, it is not clear that the vacuum angle in the hypercolor sector should be small. If $\theta_{HC}\sim{\cal O}(1)$, some phenomenology sensitive to parity (such as mixing of new states with the SM Higgs) may be qualitatively affected, while other phenomena (such as total diphoton rates) should remain qualitatively similar. For our purposes it is sufficient to focus on the case of small $\bar\theta_{HC}$, but we emphasize that new strong sectors may exhibit interesting parity-violating features.} As the mass of portal states is lowered with respect to the hypercolor confinement scale, various hyperonia also become relevant, and either the $s$-wave $\eta(0^{-+})$ or $p$-wave $\chi(0^{++})$ hyperonia states can lead to observable diphoton resonances. More generally, hyperglueballs and hyperonia may be comparable in mass, leading to various mixing effects. These examples predict a range of additional resonant and nonresonant phenomena that may serve as guideposts for future searches.

\subsubsection{Hyperglueballs}

To study the phenomena associated with $\Lambda_{HC} \sim M$, consider first the limit in which $M \gg \Lambda_{HC}$. In this case the lightest states in the hypercolor sector will be hyperglueballs. The precise hyperglueball spectrum depends on the hypercolor gauge group, but is generally expected to contain a lightest glueball with $J^{PC}$ quantum numbers $0^{++}$ and an array of nearby states including  $0^{-+}$ and $2^{++}$ (see Footnote~\ref{fn:parity} for a comment on parity). Much more is known about the case $G_{HC} = SU(3)$, where lattice predictions indicate $m_{2++} \simeq 1.4 m_{0++}$ and $m_{0-+} \simeq 1.5m_{0++}$ \cite{Morningstar:1999rf} with $m_{0++} \simeq 7 \Lambda_{HC}$ \cite{Chen:2005mg}. These hyperglueballs are all in principle suitable candidates for resonant diphoton signatures.

The interactions of hyperglueballs with the Standard Model may be parametrized by perturbatively integrating out the portal fields. This yields an Euler-Heisenberg-like effective Lagrangian connecting Standard Model bosons to hypergluons, which can be organized in terms of irreps of the Lorentz group in the hypercolor sector. The leading effective operators responsible for $0^{++}$ production and decay are \cite{Juknevich:2009ji}
\begin{eqnarray} \label{eq:glueff}
\mathcal{L} \supset \frac{\alpha_{HC}}{60 M^4} {\rm tr} \, (H_{\mu \nu} H^{\mu \nu}) \left[ c_1 \alpha_1 B_{\mu \nu} B^{\mu \nu} + c_2 \alpha_2 {\rm tr} \, (W_{\mu \nu} W^{\mu \nu}) + c_3 \alpha_3 {\rm tr} \, (G_{\mu \nu} G^{\mu \nu}) \right] + \dots 
\end{eqnarray}
where $\dots$ includes combinations of Standard Model and hypercolor field strength transforming as different irreps of the Lorentz group; these are relevant for the production and decay of glueballs with other $J^{PC}$ quantum numbers. Here the coefficients $c_i (i = 1,2,3)$ depend on the Standard Model representations of the heavy portal states. Values of $c_i$ corresponding to Standard Model quantum numbers of various portal states are listed in Table \ref{tab:cis}. To make a connection with the notation of equations~\ref{eq:theta21} and \ref{eq:theta13}, note that
\begin{equation}
\tan \vartheta_{21}=\frac{c_{2}}{2c_{1}}\qquad {\rm and}\qquad \tan \vartheta_{13}=\frac{2c_{1}}{c_{3}}.
\end{equation}

\begin{table}[h]
\begin{center}
\begin{tabular}{|c|c|c|c||c|c|c|}  \hline
$SU(3)$ & $SU(2)$ & $U(1)$ & $SU(N_{HC})$ & $c_1$ & $c_2$ & $c_3$ \\ \hline
$\Box$ & $\Box$ & $Y$ & $\Box$& $6 Y^2$  & $3$ & $2$ \\ 
$\Box$ & 1 & $Y$ & $\Box$ & $3 Y^2$ & $0$ & $1$   \\ 
1 & $\Box$ & $Y$ & $\Box$ & $ 2 Y^2$ & $1$ & 0 \\
 1 & 1 & $Y$ & $\Box$& $Y^2$ & 0 & 0 \\ \hline
\end{tabular}
\end{center}
\caption{Coefficients $c_i$ in the effective Lagrangian (\ref{eq:glueff}) corresponding to vectorlike portal fermions $\Psi, \overline{\Psi}$ with various Standard Model quantum numbers. For simplicity, only the quantum numbers of $\Psi$ are shown.}
\label{tab:cis}
\end{table}%

The effective Lagrangian (\ref{eq:glueff})  leads to a partial width for the decay of $0^{++}$ glueballs into gluons of
\begin{eqnarray} \label{eq:0gg}
\Gamma(0^{++} \to gg) = \frac{\alpha_s^2}{32 \pi^3} \frac{m_0^3}{M^2} \left(\frac{c_3}{60} \frac{g_{HC}^2 f_0^S}{M^3} \right)^2
\end{eqnarray}
where $f_0^S \equiv \langle 0 | S | 0^{++} \rangle$ is the appropriate glueball matrix element. Pure $SU(3)$ lattice data gives $g_{HC}^2 f_0^S \simeq 3 m_0^3$ \cite{Chen:2005mg, Juknevich:2009gg}. Neglecting phase space factors (and the possibility of Higgs mixing), the branching ratios to other SM diboson final states are given in terms of relative ratios by
\begin{eqnarray}
\frac{\Gamma_{0^{++} \to \gamma \gamma}}{\Gamma_{0^{++} \to gg}} &=&  \frac{1}{8} \frac{\alpha^2}{\alpha_s^2} \left(\tan \vartheta_{13}\left(1+\tan \vartheta_{21}\right)\right)^{2}\\
 \frac{\Gamma_{0^{++} \to ZZ}}{\Gamma_{0^{++} \to gg}} &=& \frac{1}{8} \frac{\alpha^2}{\alpha_s^2} \left(\tan \vartheta_{13}\left(t_W^2+  t_W^{-2} \tan \vartheta_{21}\right)\right)^{2}\\
 \frac{\Gamma_{0^{++} \to Z\gamma }}{\Gamma_{0^{++} \to gg}} &=&\frac{1}{4} \frac{\alpha^2}{\alpha_s^2} \left(\tan \vartheta_{13}\left(t_W-  t_W^{-1} \tan \vartheta_{21}\right)\right)^{2}  \\
  \frac{\Gamma_{0^{++} \to WW}}{\Gamma_{0^{++} \to gg}} &=&\frac{1}{4} \frac{\alpha^2}{\alpha_s^2} \left(s_W^{-2} \tan \vartheta_{13} \tan \vartheta_{21}\right)^{2}
\end{eqnarray}
in keeping with the general expectations developed in Section \ref{sec:independent}. Here $s_W$ and $t_W$ denote the sine and the tangent of the Weinberg angle.

In terms of the diphoton excess near 750 GeV, a $0^{++}$ glueball explanation of the signal would require $m_{0++} \simeq 750$ GeV, corresponding to a confinement scale of $\Lambda_{HC} \sim 100$ GeV. For reasonable choices of portal matter representations the partial width (\ref{eq:0gg}) leads to a $\sqrt{s} = 13$ TeV resonant production cross section for a 750 GeV $0^{++}$ state on the order of
\begin{eqnarray} \label{eq:0xs}
\sigma_{LO}(pp \to 0^{++})\simeq 1 \, {\rm fb} \; \times (m_0 / M)^8
\end{eqnarray}
where we have assumed the $SU(3)$ lattice value for the glueball matrix element $f_0^S$. This is roughly three orders of magnitude too small to accommodate the observed diphoton rate even when $m_0 / M \sim 1$, once branching ratios are taken into account. The strong dependence on $m_0 / M$ rapidly worsens the situation as portal states are decoupled. This makes it unlikely that the observed resonance is due to resonant production of a hyperglueball of a theory with a confinement scale well separated from the masses of the portal states.  Although heavier glueballs such as the $0^{-+}$ and $2^{++}$ may also be produced via gluon fusion and decay into diphotons, the rates for these process are on the order of the $0^{++}$ rate in the case of the $0^{-+}$, and well below the $0^{++}$ rate in the case of the $2^{++}$.

With this in mind, there are three possible scenarios in which hyperglueballs might remain directly relevant in light of current experimental sensitivity:
\begin{enumerate}

\item The interaction between the hypercolor sector and the Standard Model gauge group is not well modeled by perturbative exchange of heavy portal states, so that the coefficients in (\ref{eq:glueff}) are increased by one or more orders of magnitude. Indeed, the suppression in (\ref{eq:0gg}) arises in large part from the perturbative coefficients in the effective Lagrangian (\ref{eq:glueff}).
\item The hypercolor sector is not well characterized by existing $SU(3)$ lattice data for glueball matrix elements, leading to substantially larger matrix elements than were used in (\ref{eq:0xs}).
\item Glueballs are produced not by resonant $s$-channel production, but by the annihilation decays of portal states with correspondingly larger pair production cross sections. Roughly speaking, the observed rate could be reproduced by $M \lesssim 700$ GeV depending on the coefficients $c_i$, in which case the physics of the portal states is itself directly relevant. Moreover, glueballs appearing in the annihilation decays of portal states would generically be accompanied by considerable activity in the final state, including additional glueballs and Standard Model radiation; this appears inconsistent with the apparent properties of the observed excess. 
\end{enumerate} 

\noindent {\bf Additional Signatures:} \\

Should a diphoton signal arise from the decays of a hyperglueball due to one of the above considerations, there are a variety of associated predictions. The rich spectrum of higher $J^{PC}$ glueballs leads to a variety of nearby resonances with gluon fusion production modes and $\gamma \gamma, Z \gamma, ZZ$, and possibly $WW$ final states. If the glueball spectrum from lattice studies of $SU(3)$ is a guide, a $0^{++}$ hyperglueball at 750 GeV would be accompanied by $0^{-+}$ and $2^{++}$ hyperglueballs around 1-1.2 TeV, with additional $J^{PC}$ resonances at higher mass. 
This assumes the diphoton signal is due to the production of a $0^{++}$ hyperglueball. If the signal is instead due to a heavier $0^{-+}$ or $2^{++}$ state, there may be (rare) additional decays of the 750 GeV state to a $0^{++}$ glueball plus pairs of Standard Model gauge bosons (typically gluons), as well as a lower-mass $0^{++}$ resonance (around 500-600 GeV if $SU(3)$ lattice data is a guide).

Whether the glueballs are produced resonantly or in the annihilation decays of portal states, the portal states are unlikely to be much heavier than the glueballs themselves. This further implies interesting signatures from annihilation decays of portal states, as well as the rich phenomenology of hyperonia.

\subsubsection{Hyperonia}

As the portal mass $M$ is lowered, bound states of portal matter -- hyperonia -- become increasingly relevant. Strong hypercolor interactions will produce spin-0 hyperonia (the analogues of familiar $\chi$ and $\eta$ quarkonia), as well as a variety of higher-spin states, and it is possible for a hyperonium state to generate a resonance at (say) 750 GeV provided $M \sim 375$ GeV. Any spin-0 SM-singlet resonance will be accompanied by a variety of other states of comparable mass with different Standard Model quantum numbers. For example, for portal matter transforming as a bifundamental under QCD and hypercolor, the corresponding hypercolor-singlet -onia will arise in both QCD-singlet and QCD-octet representations. The former may give rise to diphoton and other diboson signatures, while the latter are copiously produced and subject to more stringent constraints. 

For concreteness, in this section we focus on the case of portal matter consisting of a single pair of vectorlike fermions $\Psi, \overline \Psi$ with the $\Psi$ transforming as a bifundamental under QCD and hypercolor with hypercharge $Y_\Psi$ (equal to electric charge $Q_\Psi$) and mass $M_\Psi$. We take $\Lambda_{HC}\lesssim M_\Psi \lesssim 7\Lambda_{HC}$, so that
various hypercolor-singlet bound states of $\Psi, \overline \Psi$ are present at the bottom of the spectrum. The most tractable of these states are the $s$-wave color-singlet and octet pseudoscalar hyperonia $\eta^0_\Psi$ and $\eta^8_\Psi$, with masses $M_0$ and $M_8$ that to first approximation are given by 
\begin{align}
M_0 \sim M_8 \sim 2 M_\Psi \equiv M_\Phi\;.
\end{align}
  The $p$-wave color-singlet and octet scalar hyperonia $\chi^0_\Psi$ and $\chi^8_\Psi$ are also of considerable interest, although as $p$-wave states their properties are somewhat less tractable analytically.

Consider first the color-singlet hyperonium $\eta^0_\Psi$. This state inherits couplings to Standard Model gauge bosons that may be computed in terms of the ground state of the nonrelativistic Hamiltonian:
\begin{eqnarray}
\Gamma(\eta^0_\Psi \to gg) &=& \frac{8}{3} N_{HC }\frac{\alpha_s(M_\Phi)^2}{M_\Phi^2}|\psi(0)|^2 \\
\Gamma( \eta^0_\Psi \to \gamma\gamma) &=& 12 N_{HC} \frac{\alpha(M_\Phi)^2 Q_\Psi^4}{M_\Phi^2}|\psi(0)|^2
\end{eqnarray}  
where $\psi(0)$ is the radial ground-state wavefunction at the origin. (For clarity, we make explicit the scale at which $\alpha, \alpha_s$ are evaluated.)
 
To estimate the ground-state wavefunction at the origin, we need a parameterization of the potential. The nonrelativistic Coulombic potential describing exotic quarks bound by perturbative $SU(N_{HC})$ exchange is
\begin{align}
V(r)=-C\frac{\bar\alpha_{HC}}{r}\;.
\end{align}
Here the coupling constant is evaluated at the inverse Bohr radius, $\bar\alpha_{HC}\equiv\alpha_{HC}(a_0^{-1})$, $a_0^{-1}=\frac{1}{2}C\bar\alpha_{HC}M_\Psi$.  $C$ is a group theory factor that depends on the hypercolor representations of the $\Psi$s and the bound state, $C\equiv C_2(R_\Psi)$ for hypercolor-singlet bound states. For hypercolor-fundamental $\Psi$, $C=(N_{HC}^2-1)/2N_{HC}$. 
With the Coulomb potential, the factor $|\psi(0)|^2$ in the partial width above is
\begin{align}
|\psi(0)|^2=\frac{1}{\pi a_0^3}=\frac{1}{8\pi}(C\bar\alpha_{HC}M_\Psi)^3\;.
\end{align}
Although it does not capture the long-distance physics of confinement, this potential provides a reasonable approximation for the parametric scaling of $s$-wave wavefunctions \cite{Barger:1987xg}.

\begin{figure}[t]
\centering
\vspace*{1cm}
\includegraphics[width=0.48\textwidth]{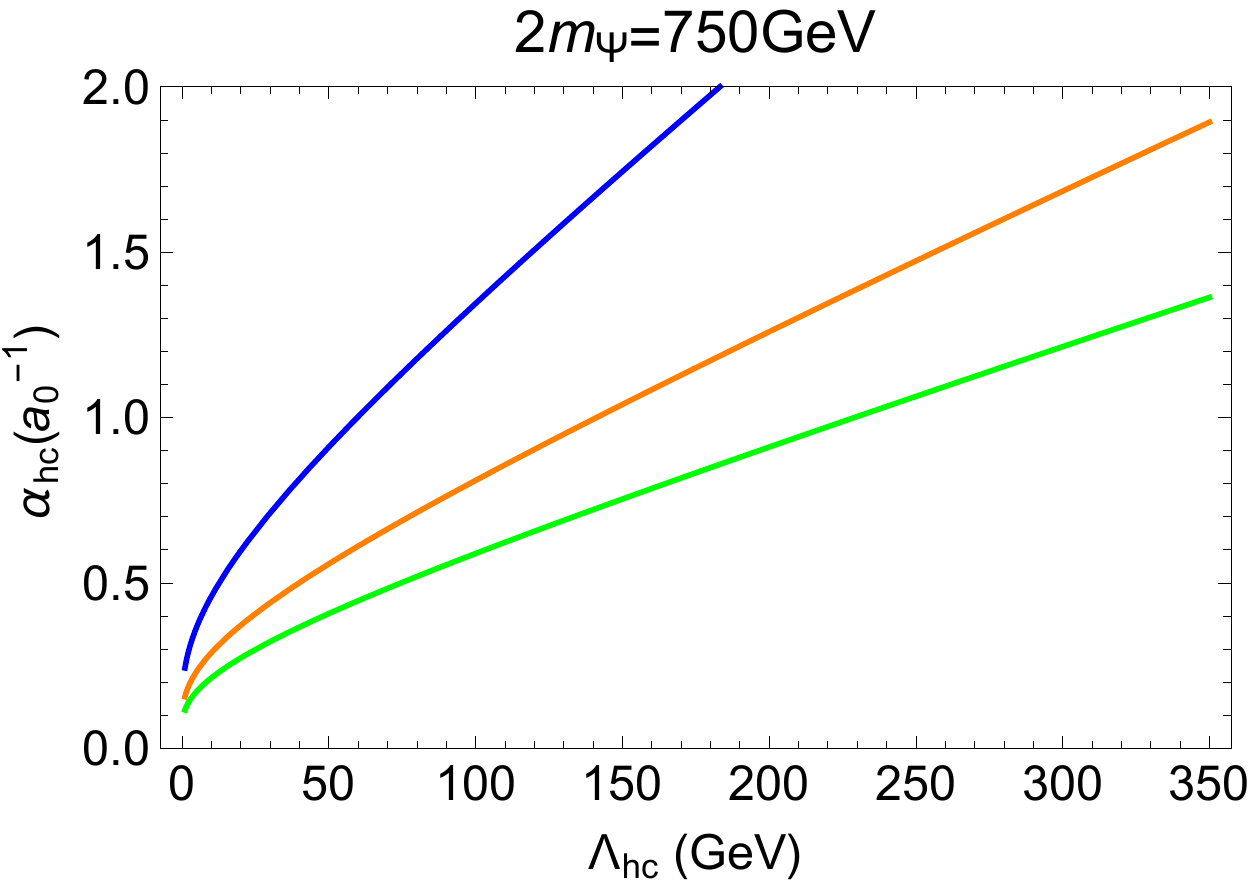}
\includegraphics[width=0.48\textwidth]{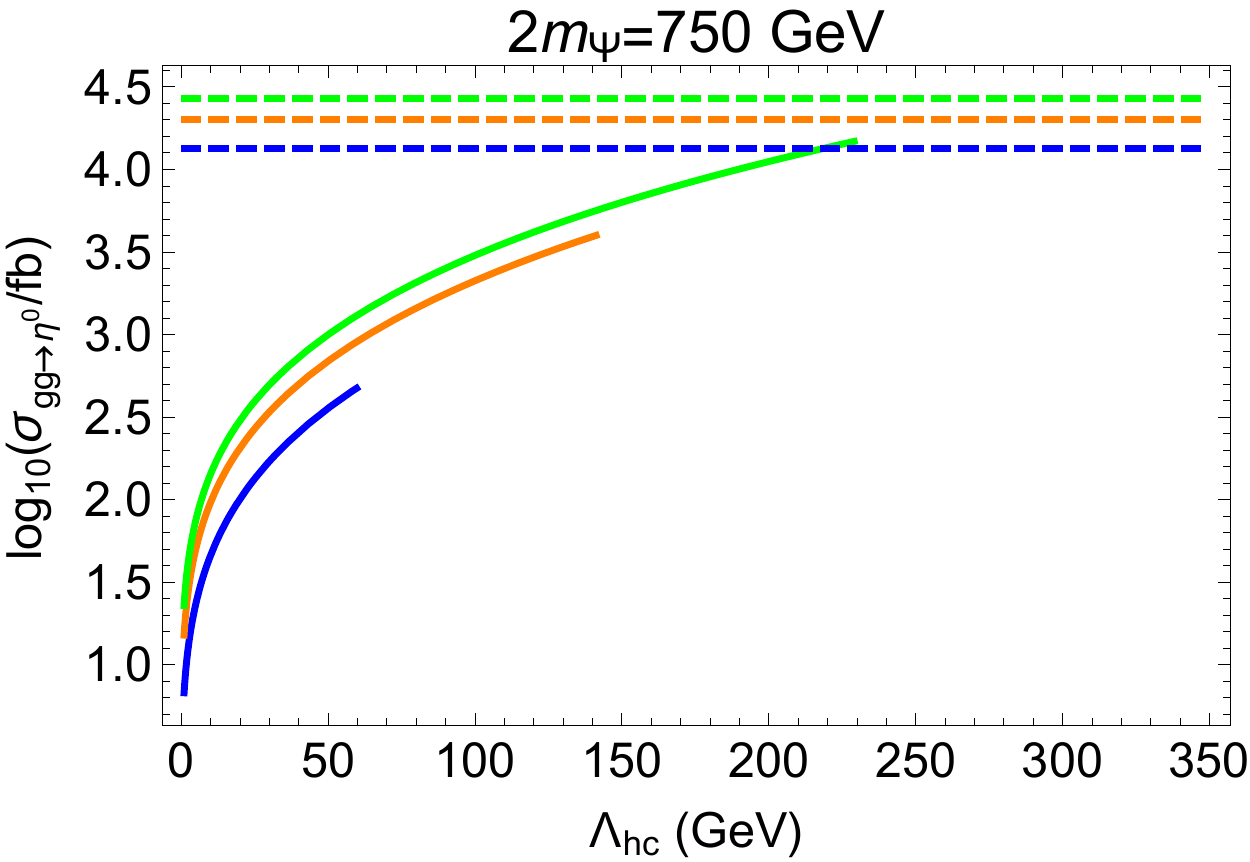} 
\caption{Left: The running hypercoupling evaluated at the Bohr radius. Right: the $gg\rightarrow \eta^0_\Psi$ production rate at 13 TeV, requiring $\bar\alpha_{HC}<1$. The dashed lines denote the QCD pair production cross section $gg\rightarrow \Psi\overline \Psi$ at 13 TeV. In both figures, curves are plotted for $N_{HC}= \{$2 (blue), 3 (orange), 4 (green)$\}$. }
\label{fig:alphaandxsec}
\end{figure}

The hyperonia parameter space relevant to the possible diphoton excess at $750$ GeV is subject to theoretical constraints. Given $M_\Phi=750$ GeV and a value for $N_{HC}$, $\Lambda_{HC}$ should be less than $M_\Psi \sim 375$ GeV to ensure that explicit chiral symmetry breaking is more important than spontaneous chiral symmetry breaking. However, perturbativity of $\bar\alpha_{HC}$ provides a stronger bound on $\Lambda_{HC}$ (Fig.~\ref{fig:alphaandxsec}a).

There are two sources of production for hyperonia: resonant production (e.g. $pp \to \eta_\Psi^0 + X$) and continuum production (e.g. $pp \to \Psi \overline \Psi + X$).\footnote{For an excellent general discussion of resonant and continuum production, see \cite{Kats:2009bv}.} The resonant production cross section is generally significantly smaller than the continuum rate, and leads to relatively clean final states. While continuum production also leads to hyperonia bound states, it proceeds in a far more chaotic fashion: $\Psi$ and $\overline \Psi$ are produced in a hypercolor-singlet state connected by a hypercolor string, which ultimately brings the $\Psi, \overline \Psi$ back together (in contrast to, e.g. open production of charm, which need not end in charmonia). The annihilation decay of the continuum $\Psi \overline \Psi$ pair can produce hyperonia that then decay into Standard Model particles, albeit with substantial additional radiation into both Standard Model final states and hypercolor final states (if hyperglueballs are light). diphoton final states are likely to be accompanied by significantly more event activity than is apparent in the observed diphoton excess. Amusingly, then, {\it diphoton events that may not pass tight isolation cuts} are an interesting prediction of annihilation decays of continuum $\Psi \overline \Psi$ pairs.

For the sake of definiteness, then, we focus on the case of resonant production, where at leading order, the singlet $\eta^0_\Psi$ has cross section
\begin{align}
\sigma(pp \to \eta_\Psi^0)=\frac{\pi^2}{8 M_\Phi^3}{\cal L}\times\Gamma(\eta^0_\Psi \to gg)\;,
\end{align}
where $\cal L$ is the gluon luminosity factor.  In Fig.~\ref{fig:alphaandxsec}b we give the 13 TeV resonant production cross section, conservatively requiring $\bar\alpha_{HC}<1$ for perturbativity. We also show the continuum cross section for $gg \to \Psi \overline \Psi$.

As for the decay of the $\eta_\Psi^0$, the ratio of branching ratios to $\gamma \gamma$ and $gg$ for the color-singlet hyperonium is 
\begin{align}
\frac{\Gamma(\eta^0_\Psi \to \gamma \gamma)}{\Gamma(\eta^0_\Psi \to gg)}=\frac{9\alpha(M_\Phi)^2Q_\Psi^4}{2\alpha_s(M_\Phi)^2}\;.
\end{align}
For $M_\Phi=750$ GeV, this is about $3\%\times Q_\Psi^4$. 
Thus the $\eta_\Psi^0$ hyperonium state is a plausible candidate for an observable diphoton signal at 750 GeV, so long as the total width is of order the partial width into gluons, and the charge $Q_\Psi$ is not too small. Of course, in addition to decays into Standard Model final states, hyperonia may decay into hyperglueballs when kinematically accessible. For a $\eta_\Psi^0$ state at 750 GeV, decays into glueball pairs are open for $\Lambda_{HC} \lesssim 50$ GeV, given that $m_{0++} \sim 7 \Lambda_{HC}$.  Preserving an appreciable diphoton branching ratio therefore likely requires $\Lambda_{HC}>50$ GeV so that decays to hyperglueballs are kinematically shut off.

It is also possible for the $p$-wave scalar $\chi_\Psi^0$ hyperonium state to give rise to a diphoton resonance. Here it is somewhat more difficult to make reliable predictions, since the Coulomb potential provides a poor approximation for (the derivative of) $p$-wave wavefunctions at the origin. Nonetheless, we may still make reliable predictions for ratios of branching ratios for which dependence on the wavefunction drops out, in particular

\begin{eqnarray}
\frac{\Gamma(\chi^0_\Psi \to \gamma \gamma)}{\Gamma(\chi^0_\Psi \to gg)}=\frac{9\alpha(M_\Phi)^2Q_\Psi^4}{2\alpha_s(M_\Phi)^2}\;.
\end{eqnarray}

It is plausible that the resonant production rate for $\chi_\Psi^0$ hyperonium at $M_\Phi = 750$ GeV is compatible with the observed excess, but without an accurate parametrization of the $p$-wave wavefunction at the origin we will not attempt a quantitative study. As in the case of the $\eta_\Psi^0$, a color-singlet $\chi_\Psi^0$ state near 750 GeV would be accompanied by a QCD-octet counterpart. \\

\noindent {\bf Additional Signatures:} \\

\begin{figure}[t]
\centering
\vspace*{1cm}
\includegraphics[width=0.48\textwidth]{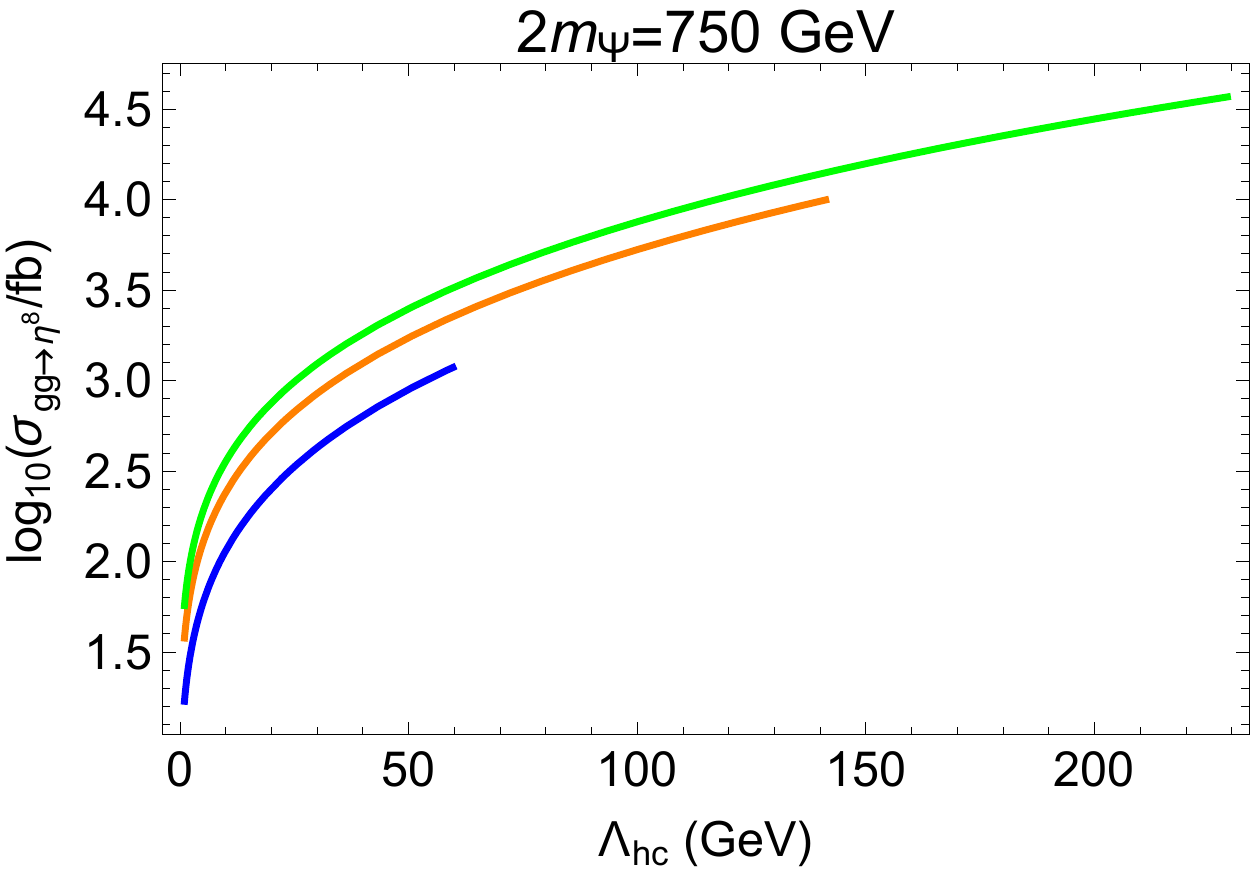} 
\caption{The $gg\rightarrow \eta^8_Q$ production cross section at 13 TeV, requiring $\bar\alpha_{HC}<1$. Curves are as in Fig.~\ref{fig:alphaandxsec}. }
\label{fig:octetxsec}
\end{figure}

A resonant diphoton signature due to hyperonium decays would be accompanied by a rich variety of associated signals. In particular, the $\chi_\Psi^0$ and $\eta_\Psi^0$ states would be accompanied by a color-octet (pseudo)scalar of comparable mass. 
The color-octet state $\eta_\Psi^8$ may be produced resonantly via its induced coupling to gluons, or in pairs via its color charge. The resonant production cross section is
\begin{align}
\sigma(pp \to \eta_\Psi^8)=\frac{\pi^2}{M_\Phi^3}{\cal L}\times\Gamma(\eta^8_\Psi \to gg)\;.
\end{align}
Interesting final states for the scalar octet include $gg$ and $g\gamma$. The ratio of branching ratios into these final states is
\begin{align}
\frac{\Gamma(\eta^8_\Psi \to g \gamma)}{\Gamma(\eta^8_\Psi \to g g)}=\frac{3\alpha(M_\Phi)\alpha_s(M_\Phi)Q_\Psi^2}{5\alpha_s(M_\Phi)^2}\;.
\end{align}
For an octet of similar mass to the singlet, around 750 GeV, this ratio is about $5\%\times Q_\Psi^2$. With this mass assumption, the leading-order (LO) resonant cross section is plotted in Fig.~\ref{fig:octetxsec}. For ${\cal O}(1)$ branching to $gg$ and most values of $\Lambda_{HC}$, the octet is not in conflict with the $\mathcal{O}({\rm pb})$-level constraints~\cite{Aad:2014aqa} from dijet searches at 8 TeV.

A host of other hyperonia of diverse $J^{PC}$ and Standard Model quantum numbers should exist in the same mass range. In particular, in analogy with quarkonium, a $J=1$ ``hyper-$J/\psi$" is expected to be relatively close in mass to the $\eta_\Psi^0$, as also noted in~\cite{Agrawal:2015dbf}. For the ordinary $J/\psi$, annihilation decays through $\gamma^*$ to dileptons comprise ${\cal O}(10)\%$ of the branching fractions, and similar rates might reasonably expected for the hyper-$J/\psi$. However, as a spin-1 color-singlet state the hyper$J/\psi$ lacks the resonant gluon fusion production channel of the $\eta_\Psi^0$; its dominant resonant production mode is $gg \to (J/\psi)^0_\Psi + g$, with a corresponding cross section some two orders of magnitude below that of $\eta_\Psi^0$. Of course, the hyper-$J/\psi$ may also be produced via continuum production, albeit with an incalculable rate. Thus we conclude that the resonant dilepton signal of a nearby hyper-$J/\psi$ is likely compatible with existing limits but provides a promising additional signal channel.

In addition to the $J = 1$ states, there are additional $J = 0$ and $J = 2$ states in the same mass range. This makes it fruitful to search for additional $\gamma \gamma,  Z\gamma, ZZ$, and possibly $WW$ resonances nearby. Continuum production of $\Psi + \overline \Psi$ should be comparable to, or greater than, the rate for resonant hyperonium production, giving rise to events with considerable additional soft radiation in addition to two or more hard Standard Model final states. 

Before we turn to the next regime, it is worth highlighting a few key differences between the hyperglueball and hyperonia cases. In the case of hyperglueballs, the resonance merely couples to the charged and colored matter. The production rate at the LHC is determined entirely by the QCD-charged portal states, but the relative branching ratios to photons and gluons are determined by the electroweak-charged portal states, which may not be identical. As discussed earlier, this allows variations in the relative branching ratios depending on the quantum numbers of the portal states. Another key feature of hyperglueballs is the lack of hyperglueball states transforming under QCD; all couplings to the Standard Model are generated by intermediary portal states. These features are in contrast to a hyperonium resonance, which is comprised of charged and colored matter; the production and decay rates for a given resonance are determined uniquely by the Standard Model quantum numbers of the portal states bound in the resonance. Moreover, assuming the portal matter is bifundamental under QCD and hypercolor, for every color-neutral hyperonium resonance there must also be a color-octet hyperonium resonance of comparable mass. More generally, it is possible for the signal to arise from mixing between hyperglueballs and hyperonia, in which case a multiplicity of states of various quantum numbers is expected near the observed resonance.

\subsubsection{Higgs mixing}

Thus far our discussion of hyperglueballs and hyperonia has neglected the possibility of mixing with the SM Higgs. Here such mixing arises if, for example, there are fermionic portal states with Yukawa couplings to the Higgs. In the case of hyperglueballs, integrating out portal states generates dimension-6 operators coupling the hypercolor sector to $|H|^2$~\cite{Juknevich:2009gg}, which for nonzero vacuum angle $\theta_{HC}$ has the form
\begin{eqnarray}
\mathcal{L} \supset \frac{\alpha_{HC} y^2}{3 \pi M^2} |H|^2 \left[{\rm tr}(H_{\mu \nu}H^{\mu\nu})+i c~\theta_{HC}{\rm tr}(H_{\mu \nu}\tilde{H}^{\mu\nu})\right]
\end{eqnarray}
where $c$ is a constant and $y$ is the Higgs-portal Yukawa coupling (up to possible $\mathcal{O}(1)$ factors depending on the quantum numbers of the portal states involved in the interaction). For small $\theta_{HC}$, this operator leads to mixing between the SM-like Higgs and the $0^{++}$ hyperglueball. The  mixing angle is of order
\begin{eqnarray}
\sin \theta_{\phi h} \sim \frac{y^2 \alpha_{HC} v f_0^S}{3 \pi m_0^2 M^2 } 
\end{eqnarray}
This leads to a variety of effects of the type discussed in Section \ref{sec:independent}, albeit with an intrinsically small mixing angle. Note that while this provides a new resonant production channel for the $0^{++}$, Higgs-glueball mixing is too small to significantly alter the resonant production cross section in (\ref{eq:0xs}).

In the case of hyperonia, Yukawa couplings can give rise to mixing between the SM-like Higgs and the $\chi_\Psi^0$. Assuming there are several portal states with appropriate quantum numbers and comparable mass, the Higgs-onium mixing is then of order \cite{Drees:1989du}
\begin{eqnarray}
\sin \theta_{\phi h} \sim \frac{y |\psi^\prime(0)|}{\sqrt{\pi }M^{5/2}} 
\end{eqnarray}
where $\psi^\prime(0)$ is the derivative of the $p$-wave radial wavefunction at the origin.

\subsection{Hyperpions}

Finally, consider the limit $M\ll\Lambda_{HC}$, in which most hypercolor bound states obtain the dominant contribution to their masses from confinement. If hypercolor confinement spontaneously breaks global symmetries of the theory, there will be parametrically light pionlike states with interesting resonant production and decay modes at the LHC.

The behavior of hypercolor theories with vectorlike hyperquarks ($M\ll\Lambda_{HC}$), dubbed ``Vectorlike Confinement'' (VC) theories, was studied in detail in ref.~\cite{VC} and details of the phenomenology of certain VC models were further explored in refs.~\cite{coloron,coloronLHC,VC-pheno,Bai:2010qg,Antipin:2015xia}. VC models have already been considered as the source of the diphoton resonance~\cite{Franceschini:2015kwy, Harigaya:2015ezk,Nakai:2015ptz,Molinaro:2015cwg,Matsuzaki:2015che,Bian:2015kjt,Bai:2015nbs,Cline:2015msi}. Below we will briefly summarize the salient features in VC theories that are most relevant for the connection to the diphoton resonance. We will focus on the $\SU(N_{HC})$ as the hypercolor group. The matter content in the UV is composed of a number of vectorlike hyperquarks $\psi$ that have Dirac masses and that transform under both hypercolor and the SM gauge interactions. We will refer to each such fermion bilinear as a ``species''. In the absence of higher-dimensional operators, there is an unbroken global $\U(1)$ symmetry associated with each species number.

The lightest mesons in the spectrum are the pseudoscalars (``hyperpions'' $\hpi$)\footnote{For $N_{F}$ hyperflavors, we define $\tilde{\pi}=\tilde{\pi}^{a}T^{a}$ with the normalization ${\rm Tr}[T^{a}T^{b}]=\delta^{ab}/2$.}. These fall into two categories, those with zero and nonzero species number, referred to as \hpiSx and as \hpiL, respectively. Depending on the matter content, one or more \hpiLx are rendered stable by the conserved species numbers, and these can decay only if higher-dimensional operators are added to the UV theory, which may be suppressed by a high mass scale ($\gg \Lambda_{HC}$). The \hpiSx on the other hand can decay through the axial anomaly to any pair of SM gauge bosons under which the constituent hyperquarks are charged. This is encoded in the effective Lagrangian by the term
\begin{equation}
{\mathcal L}\supset \frac{N_{HC}\epsilon^{\mu\nu\alpha\beta}}{16\pi^{2}\fpi}\sum_{i,j} g_{i} g_{j} {\rm Tr}\left[\tilde{\pi} F_{i,\mu\nu} F_{j,\alpha\beta} \right]
\label{eq:anomaly}
\end{equation}
where $i,j$ run over the SM gauge groups and in the large-$N$ limit $f_{\tilde{\pi}}\approx \sqrt{N_{HC}}\  \Lambda / 4\pi$. Here we have introduced the notation $\Lambda$ which is the equivalent of $\Lambda_{\rm QCD}\approx 1~$GeV for hypercolor. $\Lambda$ should not be confused with the confinement scale $\Lambda_{\rm HC}$, as they differ by a multiplicative factor.

Hyperpions get masses from three sources:
\begin{itemize}
\item From hyperquark masses, $m_{\tilde{\pi}}^{2}\approx M_{\psi} \Lambda$. This contribution exists for all hyperpions\footnote{For clarity, throughout this section, ``quark mass'' will always mean constituent quark mass}.
\item Hyperpions that are not singlets under the SM gauge groups receive an additional mass correction
\begin{equation}
\delta m_{\tilde{\pi}}^{2} \approx \frac{3\Lambda^{2}}{16\pi^{2}}\sum g_{i}^{2} C_{2}^{i}(\tilde{\pi})
\label{eq:gaugemass}
\end{equation} 
where the sum is over all SM gauge groups that $\tilde{\pi}$ is charged under, and the $C_{2}^{i}$ are the corresponding Casimirs in the appropriate representation. From this contribution alone, $\tilde{\pi}$ that are QCD color octets receive a mass of $\approx 0.3 \Lambda)$.
\item Hyperpions that are not singlets under $\SU(2)_{L}$, receive additional ``hyperfine'' mass splittings from electroweak symmetry breaking, which are calculable but will not be important in this study.
\end{itemize}

The $\tilde{\pi}$-long cannot be resonantly produced but they may be pair produced if they carry SM charges. The $\tilde{\pi}$-short on the other hand can be resonantly produced through the anomaly vertex of equation~\ref{eq:anomaly}. Even though the vertex in question is loop-suppressed, if the hyperpions are sufficiently heavy, then the resonantly produced \hpiSx will be the first states that are observed. The branching fractions of these states can be calculated using (\ref{eq:anomaly}).

\subsubsection{Benchmark model}

Let us now construct a minimal benchmark model that can explain the diphoton resonance. Since the observed hyperpion must couple to $\gamma$-$\gamma$, it must be a SM gauge singlet \hpiS. In order for it to be produced with appreciable cross section, it must also couple to $g$-$g$, and therefore the constituent hyperquarks must carry SM color as well as hypercharge. For simplicity, we will take the hyperquarks to be $\SU(2)_{L}$ singlets. Let us label the diphoton resonance as $\tilde{\pi}^{0}$.

Note that having a single hyperquark species is not a viable option, because the only SM-singlet hyperpion in this case corresponds to the $\tilde{\eta}'$, which has a mass of the order of $\Lambda$. As we will see below, not only would this signal the existence of even lighter hyperpions that are experimentally excluded, but we need the $m_{\tilde{\pi}^{0}}$ to be lighter than $\Lambda$ in order to reproduce the correct production cross section. Thus the minimal model must contain two hyperquarks, at least one of which is colored. If the second hyperquark is taken to be a color singlet, then the lightest \hpiLx will be colored, and a higher-dimensional operator needs to be added in the UV to allow it to decay. We will pursue a different choice such that the lightest \hpiLx is a SM gauge singlet, and we will not explicitly break the conserved species number symmetry that keeps it stable. The charge assignment for the benchmark model is presented below.
\begin{table}[h]
\begin{center}
\begin{tabular}{c|c||c|c|c}
  & $\SU(N)_{\rm HC}$ & $\SU(3)_{\rm C}$ & $\SU(2)_{\rm W}$ & $\U(1)_{\rm Y}$  \\
\hline
$\psi_{1}$ & $\Box$ & $\Box$    & ${\bf 1}$ & $q_{Y}$  \\
$\psi_{1}^{c}$ & $\overline{\Box}$ & $\overline{\Box}$ & ${\bf 1}$ & $-q_{Y}$\\
$\psi_{2}$ & $\Box$ & $\Box$    & ${\bf 1}$ & $q_{Y}$  \\
$\psi_{2}^{c}$ & $\overline{\Box}$ & $\overline{\Box}$ & ${\bf 1}$ & $-q_{Y}$
\end{tabular}
\end{center}
\caption{The quantum numbers of the hyperquarks in the benchmark model.}
\label{tab:charges}
\end{table}
With this choice, the QCD coupling still runs to zero in the UV, however for $q_{Y}=1$, $\U(1)_{Y}$ develops a Landau pole around $10^{11}~$GeV. While this is a potential worry, it is possible that hypercharge is unified into a non-Abelian group at an intermediate scale which is asymptotically free.

In accordance with equation~\ref{eq:anomaly}, the branching ratio $\Gamma(\tilde{\pi}^{0}\rightarrow \gamma\gamma)/\Gamma(\tilde{\pi}^{0}\rightarrow gg)$ can be calculated to be $0.033\, q_{Y}^{4}$ (assuming $q_{Y}\lsim 1$). For the benchmark model we will adopt the choices $q_{Y}=1$ and $N_{HC}=3$. Since the cross section $gg\rightarrow \tilde{\pi}^{0} \rightarrow \gamma\gamma$ measured by ATLAS and CMS is 5~fb, this sets the total production cross section of $\hpiN$ to be 150~fb. Note that this is significantly below the dijet resonance bound at this mass. Next, we will use the production cross section to determine $\fpi$.

Note that if the hyperquarks are mass degenerate, then the generator for the $\tilde{\eta}'$ is proportional to the identity in flavor space, whereas the generator for $\hpiN$ is proportional to $\sigma^{3}$. However, since the generators of QCD and electromagnetism are also proportional to the identity in flavor space, in this limit the $\hpiN$ does not couple to the anomaly, and therefore cannot be produced resonantly. In fact, the only way for the $\hpiN$ to be produced is through mixing with the $\tilde{\eta}'$, which requires an explicit breaking of the $U(2)$ flavor symmetry by nondegenerate hyperquark masses $M_{\psi_{1}}\neq M_{\psi_{2}}$.

In the $\tilde{\eta}'$-$\hpiN$ sector, the model has four relevant parameters, $M_{\psi_{1}}$, $M_{\psi_{2}}$, $\fpi$ (or equivalently $\Lambda$), and an additional contribution to the $\tilde{\eta}'$ mass from the anomaly. There are three constraints on these parameters, the lighter mass eigenvalue, which we fix at 750~GeV, the production cross section for the 750~GeV mass eigenstate which we fix at 150~fb as discussed above, and finally the heavier mass eigenvalue which we will take to scale as $\sqrt{N_{HF} / N_{HC}} \Lambda$, which is the naive scaling for the $\tilde{\eta}'$ mass. Therefore there is a one-parameter family of solutions that satisfy all constraints. As a benchmark, we will choose the parameter point such that max$(M_{\psi_{1}} / \Lambda , M_{\psi_{2}} / \Lambda)$ is as small as possible. In particular, we will choose $\fpi=$275~GeV ($\Lambda=$2~TeV), $M_{\psi_{1}}=$400~GeV, $M_{\psi_{2}}=$1.3~TeV\footnote{In order to have an apples-to-apples comparison, we point out that using the same normalization for quark masses, the strange quark mass in QCD would be 471~MeV, where we take $\Lambda_{\rm QCD}=1~$GeV. Thus it is not a significantly worse approximation to consider $\psi_2$ as a ``light'' quark than it is to do the same for the strange quark.}.

Note that even at this optimized point one of the hyperquarks is fairly heavy. This appears to be a weak point of the benchmark model we have chosen, however it should not be taken as a weakness of the hyperpion scenario in general. A choice of hyperquark representations different than the one in Table~\ref{tab:charges} should lead to models where all hyperquarks can be chosen relatively light and still reproduce the correct production cross section for the 750~GeV mode. However, even in our minimal benchmark model there is a rich phenomenology as we will outline below, and to keep the discussion as simple as possible we will continue with this benchmark model to highlight the generic features of the hyperpion scenario. The reader should be aware that corrections to the chiral Lagrangian that are proportional to the heavy quark mass may change our estimates in the discussion of the phenomenology below by order one factors.\\

\noindent {\bf Additional Signatures:}\\

Having defined our benchmark model, let us now study the spectrum, focusing in particular on the collider phenomenology of various states. There are two \hpiS s and a complex \hpiLx that are complete SM gauge singlets. From the singlet $\tilde{\pi}$-shorts, one of them is the diphoton resonance $\tilde{\pi}^{0}$ and the other is the $\tilde{\eta}'$. The singlet complex \hpiLx (referred to as $\hpi_{L}^{1}$ from here on) is of little phenomenological interest at the LHC, since it has no couplings to the SM gauge bosons or fermions. Since it is stable however, it can be a dark matter (DM) candidate. We will come back to this point below. The model also contains two real and one complex color-octet hyperpions, the former being \hpiS s (both referred to as $\hpi_{S}^{8}$ from here on) and the latter being an \hpiLx (referred to as $\hpi_{L}^{8}$ from here on).\\

{\bf Other final states connected to the $\hpiN$:} Any particle that couples to $\gamma$-$\gamma$ through a $B_{\mu\nu}\tilde{B}^{\mu\nu}$ term will also inherit couplings to $\gamma$-$Z$ and $Z$-$Z$, and this is true for the $\hpiN$ as well. Given the leptonic branching ratios of the $Z$ however, it will take some time until these modes can be observed with statistical significance. The branching fractions to to $\gamma$-$\gamma$, $\gamma$-$Z$ and $Z$-$Z$ are universal, and are consistent with those calculated in section~\ref{sec:independent}. Of course, the dominant decay mode of the $\hpiN$ is $g$-$g$, and even though the 150~fb production cross section is currently below the dijet resonance bounds, this is a robust associated signature that should be looked for in the future. Being a gauge singlet, the $\hpiN$ cannot be pair produced through SM processes.\\

{\bf The $\tilde{\eta}'$:}  At our benchmark point, the $\tilde{\eta}'$ has a mass around 3~TeV. In terms of its production and decays, it is essentially as a heavier copy of the $\hpiN$. Due to its large mass however, its production cross section is only of order 10~fb and therefore it would require a high luminosity to look for the dominant decay into dijets, and the rare $\gamma$-$\gamma$ decay mode will almost certainly not be observable at the LHC.\\

{\bf A color-octet $\hpi$ resonance:} The color-octet $\hpi$'s receive a mass contributions both from hyperquark masses as well as due to their gauge charges in accordance with equation~\ref{eq:gaugemass}. At the benchmark point this results in color-octet \hpiSx masses of 1.1 and 1.7~TeV. With this mass, the lighter color-octet $\hpi_{S}^{8}$ has a resonant production cross section of $\sim$1~pb at 13~TeV. This is more than an order of magnitude below the current dijet bounds, however these resonances may be accessible with high luminosity. Their existence would be an important cross-check for the VC scenario. Note that while the dominant decay mode of the $\hpi_{S}^{8}$ is into a pair of gluons, they can also decay to $g$-$\gamma$ and $g$-$Z$ in accordance with equation~\ref{eq:anomaly}. The branching fractions for these modes is down by a factor of $\alpha_{Y} / \alpha_{s}$, however the backgrounds for these final states are also smaller and therefore these decay modes may be observable not too long after the dijet mode, providing another important cross-check. Unlike the $\hpi_{S}^{8}$, the $\hpi_{L}^{8}$ cannot be resonantly produced, but all color-octets can of course be pair produced through QCD, and we turn our attention to the pair production of color-octet $\hpi$ next.\\

{\bf Pair production of color-octets:} Unlike the resonant production vertex which is loop suppressed, the pair production proceeds through ${\mathcal O}(1)$ SM gauge couplings. Since the dominant decay mode of color-octet \hpiS s is into dijets, this results in a paired dijet final state. At 13~TeV, the pair production cross section of a real 1.1~TeV color-octet scalar is ${\mathcal O}$(10~fb). While below the existing bounds for paired dijets~\cite{Khachatryan:2014lpa} based on 8~TeV collisions, this final state should be observable with a modest amount of luminosity at 13~TeV.

While the $\hpi_{S}^{8}$ decay to a pair of jets, the fate of the $\hpi_{L}^{8}$ is more interesting. Like the $\hpi_{L}^{1}$, the $\hpi_{L}^{8}$ is also charged under the unbroken $\U(1)_{\psi_{1}-\psi_{2}}$. While the singlet states are potential DM candidates, stable colored particles with masses of ${\mathcal O}$(TeV) is not phenomenologically viable~\cite{Smith:1979rz,Hemmick:1989ns,Verkerk:1991jf,Yamagata:1993jq,Smith:1982qu}. Fortunately, since the $\hpi_{L}^{8}$ is heavier, it can decay to the $\hpi_{L}^{1}$, thus preserving the conserved species number, plus additional states. The leading terms that give rise to such a decay mode are contained at ${\mathcal O}(\hpi^{4})$ in the kinetic term of the chiral Lagrangian
\beq
\fpi^{2}\,{\rm Tr}\left[\partial_{\mu}\Sigma^{\dag}\partial^{\mu}\Sigma \right]\quad {\rm with} \quad \Sigma={\rm Exp}\left(i\hpi / \fpi\right).
\eeq
Specifically, through such terms the $\hpi_{L}^{8}$ can decay to the $\hpi_{L}^{1}$ and two \hpiS s, either a pair of $\hpi_{S}^{8}$ or one $\hpi_{S}^{8}$ and the $\hpiN$. At the benchmark point, the $\hpi_{L}^{8}$ mass is 1.4~TeV. Therefore, both \hpiS s in the final state need to be off shell, and they decay to $g$-$g$ through the anomaly, resulting in a 5-body decay mode ($\hpi_{L}^{1}gggg$). A very rough estimate for the width of this decay mode gives
\beq
\Gamma\left(\hpi_{L}^{8}\rightarrow\hpi_{L}^{1}(\hpiN\rightarrow gg)(\hpi_{S}^{8*}\rightarrow gg)\right)=\frac{1}{8\pi(16\pi^{2})^{3}}\left(\frac{1}{\fpi^{2}}\left(\frac{\alpha_{s}}{4\pi\fpi}\right)^{2}\right)^{2}m_{\hpi}^{9},
\eeq
where $m_{\hpi}\approx 1.4~$TeV. While this estimate neglects a number of ${\mathcal O}(1)$ factors, it should give the correct order of magnitude, and we find in fact that the decay is prompt on collider time scales. Thus, exotic final states such as R-hadrons do not appear.\\

{\bf A dark matter candidate:} As discussed above, the $\hpi_{L}^{1}$ is a stable SM gauge singlet particle, and therefore a DM candidate. At the benchmark point, it has a mass of 850~GeV. Its annihilation cross section however is too small in the benchmark model. The main annihilation channel is to $\hpiN$-$\hpiN$ through four-$\hpi$ interactions in the chiral Lagrangian. Again, a very crude estimate gives
\beq
\left(\sigma v\right)\left(\hpi_{L}^{1}\hpi_{L}^{1}\rightarrow\hpiN\hpiN\right)\approx \frac{1}{64\pi^{2}m_{\hpiN}^{2}}\left(\frac{m_{\hpiN}^{2}}{\fpi^{2}}\right)^{2}\ll\left(\sigma v\right)_{\rm thermal}.
\eeq
In a less minimal model, the cross section may be enhanced by adding other annihilation channels, for example by gauging the $\U(1)_{\psi_{1}-\psi_{2}}$ flavor symmetry, or by introducing a hidden sector.\\

{\bf Hypermesons other than the $\hpi$:} Close to the scale $\Lambda$, the theory also contains the hyper-rho mesons ($\hrho$). As discussed in ref.~\cite{VC}, for each SM gauge group that the hyperquarks are charged under, there is a $\hrho$ state that mixes with that gauge boson, with a mixing angle $\epsilon=g / g_{\hrho}$ where $g_{\hrho}$ is the strong coupling between the $\hrho$ and the $\hpi$. Unlike the $\hpi$ therefore, the $\hrho$ have direct (albeit small) couplings to the SM fermions through mixing with the SM gauge bosons. They can be resonantly produced from a $q$-$\bar{q}$ initial state at colliders\footnote{One can in principle also include higher-dimensional operators in the effective Lagrangian involving the $\hrho$, for instance leading to the production of the $g'$ from a $g$-$g$ initial state. Note however that unlike the mixing between $\hrho$ and the SM gauge bosons which sets the resonant production cross section from a $q$-$\bar{q}$ initial state and which can be estimated by scaling the $\rho$-$\gamma$ mixing in the real world, the coefficient of such higher-dimensional operators cannot be estimated from real-world data and are therefore subject to $O(1)$ uncertainties. This point is made in more detail in ref.~\cite{coloronLHC} where in fact the production of a $g'$ from a $g$-$g$ initial state was considered.}, and they dominantly decay to $2\hpi$. In our benchmark model, there is a color-octet $\hrho$ that mixes with the gluon (a $g'$) and a singlet $\hrho$ that mixes with $\gamma / Z$ (a $Z'$). At the benchmark point, they both have masses around 2~TeV, the dominant decay mode of the $Z'$ being $\hpi_{L}^{1}$-$\hpi_{L}^{1*}$ (an invisible $Z'$) and the dominant decay mode of the $g'$ being $\hpi_{S}^{8}$-$\hpiN$ (a four-jet final state, where two dijet pairs and the entire four-jet state all have a resonance peak). Note that since $q$-$\bar{q}$ is not the dominant decay mode of either $\hrho$, dijet constraints are significantly weakened. The $Z'$ also contributes to the annihilation of the DM candidate, however since the $Z'$ is spin-1 while the DM candidate is spin-0, this annihilation mode is velocity suppressed and cannot produce the cross section necessary for a thermal WIMP. The $g'$ can also be pair produced, leading to multijet final states with an even richer resonance structure.

The hypercolor sector also contains scalars (most notably, the $\sigma$ particle). In the minimal implementation of the vectorlike confinement scenario, the hyperquarks only interact with the SM through the SM gauge bosons, and therefore the scalars only mix with the Higgs boson at higher loop orders. This makes it unlikely for their resonant production to have an observable cross section. Scalars with SM charges on the other hand can of course be pair produced, but unlike the pseudoscalar $\hpi$, there is no good reason for them to be light compared to $\Lambda$.\\

{\bf Hyperbaryons:} Finally, VC theories contain hyperbaryons. The quantum numbers (including spin) of the hyperbaryons depend on $N_{HC}$ and therefore their properties are more model dependent. Generically, to avoid phenomenological problems the lightest hyperbaryon must either be a color-singlet and electrically neutral, or hyperbaryon-number violating operators need to be added to the theory in order to avoid constraints from stable charged/colored particles. For our benchmark model with $N_{\rm HC}=3$, the lightest hyperbaryons carry electric charge $\pm3$. They can be made to decay through the following higher-dimensional operator
\beq
\frac{1}{M_{UV}^{5}}\left(\psi^{c}e^{c}\right)\left(\psi^{c}e^{c}\right)\left(\psi^{c}e^{c}\right)
\label{eq:hyperbaryondecay}
\eeq
where the three factors in the brackets are contracted with a fully antisymmetric tensor in hypercolor, and in QCD. At first sight this operator appears to violate lepton number, however this can be avoided if the hyperquarks are assigned lepton number as well. Of course, if one were to add additional higher-dimensional operators to the theory, there may no longer be a way to assign lepton number consistently, leading to the breaking of lepton-number. No such operators are required for the benchmark model, and therefore we will not consider the possibility of constraints arising from lepton-number violation any further in this paper. Note also that the flavor of the leptons appearing in the operator of equation~\ref{eq:hyperbaryondecay} can be chosen such that only one flavor appears, suppressing lepton-flavor violating effects. Another interesting way to suppress lepton-flavor violation is to contract the lepton flavors with an antisymmetric tensor, thus making this operator compatible with minimal flavor violation~\cite{Chivukula:1987py,Hall:1990ac,Buras:2000dm,D'Ambrosio:2002ex,Buras:2003jf}, however in that case the flavor indices on the hyperquarks also need to be antisymmetrized for the operator to not vanish identically. This can in principle be done by adding a third hyperquark, which would further expand the spectrum of $\hpi$ states. Finally, note that depending on the hyperquark flavor structure of the operator of equation~\ref{eq:hyperbaryondecay}, individual hyperquark numbers may be broken as well, however this does not necessarily affect the stability of the DM candidate \hpiL.

In the low energy theory, this operator becomes
\beq
\frac{\Lambda^{3}}{M_{UV}^{5}}\lambda_{ijk}\tilde{\Delta}^{*}e_{i}^{c}e_{j}^{c}e_{k}^{c}
\eeq
with $\tilde{\Delta}^{*}$ being the hyper(anti)baryon with electric charge -3. The decay width can be estimated very roughly as
\beq
\Gamma\left(\tilde{\Delta}^{*}\rightarrow eee\right)\approx\frac{\lambda^{2}}{8\pi(16\pi^{2})}\frac{\Lambda^{11}}{M_{UV}^{10}}.
\eeq
Since $M_{UV}$ appears with such a high power, it cannot be too far above $\Lambda$, otherwise the hyperbaryon lifetime exceeds 1s, thus causing potential problems for early universe cosmology.\\

{\bf Future Prospects:} For $\Lambda\approx 2~$TeV, many of the states discussed above should be accessible to the LHC at high luminosity. In particular, in order of increasing energy scale, one would have a chance to study the resonant production of the color-octet \hpiS s, the resonant production of the $g'$ and $Z'$, the pair production of the color-octet \hpiSx and \hpiL, and the resonant production of the $\tilde{\eta}'$. On the other hand, the pair production of scalar or vector mesons, as well as hyperbaryons may be difficult to observe at the LHC but should be within the reach of a future $p$-$p$ collider. At such a collider, at sufficiently high energies even the formation of hyperjets should be observable.

\subsubsection{Generic predictions}

Above, we have explored the phenomenology of one benchmark model in detail. Now let us take a step back and ask which of these signatures are common to all VC models that can incorporate the diphoton resonance, and are not specific to the benchmark model\footnote{With only the assumption that the hypercolor gauge group is $\SU(N_{HC})$}. Since the resonance is produced from a $g$-$g$ initial state and it decays to $\gamma$-$\gamma$, the theory must contain at least one hyperquark that is colored and at least one (possibly the same) hyperquark that carries electroweak quantum numbers. Furthermore, as argued in the previous section, there must be at least two hyperquarks, since with only one (colored and hypercharged) hyperquark, the only $\hpi$ that has the right quantum numbers to be the diphoton resonance is the $\eta'$, and there is at least one lighter colored \hpiSx that can be resonantly produced and that would have already been observed.

Even with this minimal amount of information, we can make a number of robust predictions. The diphoton resonance must also have $\gamma$-$Z$ and $Z$-$Z$ decay modes, and optionally even a $W^{+}$-$W^{-}$ decay mode if there is at least one hyperquark that transforms under $\SU(2)_{L}$. Its dominant decay mode into $g$-$g$ robustly predicts a dijet resonance at the same mass, which is currently below the dijet resonance bounds but may be observed in the future. There must also exist at least one colored \hpiSx resonance that decays predominantly to dijets, and can therefore be looked for as a dijet resonance. This state will also have $g$-$\gamma$ and $g$-$Z$ decay modes unless it is composed entirely of hyperquarks that are colored but are electroweak singlets,.

There must also exist at least one \hpiLx which is colored. This state can be pair produced but not resonantly produced. If it is the only \hpiLx state, then it must decay to SM particles through a higher-dimensional operator, which can result in displaced vertices, or even R-hadrons that are stable on collider timescales. If there are lighter \hpiLx states, then the heavier ones can cascade decay to the lighter ones, and the lightest state may still be collider stable. It can also be a SM singlet as in our benchmark model and therefore give rise to missing energy signatures in the decay of the heavier \hpiL s.

The theory will still contain an $\eta'$, and hyper-rho mesons that have the quantum numbers of a $g'$ and $Z'$, and optionally $W'$ if there are hyperquarks that transform under $\SU(2)_{L}$. The hyper-rho mesons can be resonantly produced, and will decay to a pair of $\hpi$, which then subsequently decay, resulting in four or more SM particles. Finally, the theory will contain hyperbaryons which, if the lightest one is uncolored and electrically neutral, may be stable and contribute as a DM component.

\section{Conclusions} \label{sec:conclusions}

In this work we have illustrated experimental strategies for uncovering new physics associated with digauge boson resonances well above the weak scale. We have examined the model-independent implications of resonances that couple to Standard Model bosons through new charged and colored matter, identifying a robust set of predictions for ratios of branching ratios that can be used to determine the quantum numbers of the new states. Predictions for ratios involving one or more massless Standard Model gauge bosons are robust against mixing between the resonance and the Standard Model Higgs, while ratios involving two massive gauge bosons provide a sensitive probe of mixing. This provides a concrete set of expectations and objectives for the experimental study of new diboson resonances.

A new diboson resonance is likely to be accompanied by additional states, whether in the form of weakly coupled charged matter or additional strongly coupled resonances. To illustrate the experimental opportunities for additional states accompanying a diboson resonance, we formulate a simple universality class of theories involving a (possibly decoupled) scalar state, charged and colored portal matter, and an additional non-Abelian gauge group. Various limits of this framework give rise to observable diboson resonances, including the weakly coupled limit in which the resonance is an elementary scalar and various strongly coupled limits in which the resonance may be a glueball, quarkonium, or pion of the new gauge sector. In each case the resonance is accompanied by a variety of additional states that provide motivated targets for further experimental searches. Depending on the region of parameter space, these states include weakly coupled matter accessible in ongoing searches for partner particles; new color-octet resonances at or above the TeV scale; and new color-singlet resonances decaying into diboson and possibly difermion final states. When the resonance is a pion of the new gauge sector, confining physics also furnishes stable neutral particles that may serve as dark matter candidates.

Our results are of general relevance to any diboson resonance near the TeV scale, but are particularly timely in light of a possible excess in the diphoton spectrum near 750 GeV. The experimental strategies detailed here should be of use in exploring possible new physics implications of a genuine excess, both in existing 8 TeV and 13 TeV data and throughout the remainder of LHC Run 2. Looking to the future, the discovery of a new diphoton resonance well above the weak scale would certainly be a case of ``who ordered that?''  It is an interesting question as to whether and how such a new resonance could fit into existing paradigms for physics beyond the Standard Model.

\section*{Acknowledgements}
We would like to thank Prateek Agrawal, John Paul Chou, Jacques Distler, Yuri Gershtein, Amitabh Lath, Matthew McCullough, Sunil Somalwar, and Raman Sundrum for useful conversations. The work of NC is supported by the US Department of Energy under the grant DE-SC0014129. The research of CK is supported by the National Science Foundation under Grants No. PHY-1315983 and No. PHY-1316033. The work of ST is supported by the US Department of Energy under grant DOE-SC0010008.


\bibliography{bumprefs}
\bibliographystyle{utphys}

\end{document}